\documentclass{IEEEtran} 
\usepackage{cite}
\usepackage{amsmath,amssymb,amsfonts}
\usepackage{algorithmic}
\usepackage{graphicx}
\usepackage{textcomp}
\usepackage[hidelinks]{hyperref}

\def\BibTeX{{\rm B\kern-.05em{\sc i\kern-.025em b}\kern-.08em
    T\kern-.1667em\lower.7ex\hbox{E}\kern-.125emX}}
\begin{document}
%\history{Date of publication xxxx 00, 0000, date of current version xxxx 00, 0000.}
%\doi{10.1109/ACCESS.2017.DOI}

\title{''I think I Discovered a Military Base in the Middle of the Ocean'' -- Null Island, the Most Real of Fictional Places}
\author{\uppercase{Levente Juhász} \& \uppercase{Peter Mooney}}
%\address[1]{Geographic Information Systems Center, Florida International University, Miami, FL 33199, USA}
%\address[2]{Department of Computer Science, Maynooth University, W23 F2H6 Maynooth, Co. Kildare, Ireland}
%\tfootnote{This work received no external funding.}

\markboth
{Juhász \& Mooney: Null Island, the most real of fictional places}
{Juhász \& Mooney: Null Island, the most real of fictional places}

%\corresp{Corresponding author: Levente Juhász (e-mail: ljuhasz@fiu.edu).}

%\begin{keywords}
%error, fictional place, geocoding, geographic information science, geoweb, human-computer interaction, web mapping
%\end{keywords}

%\titlepgskip=-15pt

\maketitle

\begin{abstract}
This paper explores Null Island, a fictional place located at 0° latitude and 0° longitude in the WGS84 (World Geodetic System 1984) geographic coordinate system. Null Island is erroneously associated with large amounts of geographic data in a wide variety of location-based services, place databases, social media and web-based maps. Whereas it was originally considered a joke within the geospatial community, this article will demonstrate implications of its existence, both technological and social in nature, promoting Null Island as a fundamental issue of geographic information that requires more widespread awareness. The article summarizes error sources that lead to data being associated with Null Island. We identify four evolutionary phases which help explain how this fictional place evolved and established itself as an entity reaching beyond the geospatial profession to the point of being discovered by the visual arts and the general population. After providing an accurate account of data that can be found at (0, 0), geospatial, technological and social implications of Null Island are discussed. Guidelines to avoid misplacing data to Null Island are provided. Since data will likely continue to appear at this location, our contribution is aimed at academics, computing professionals and the general population to promote awareness of this error source.
\end{abstract}

\section{Introduction and motivation} \label{sec:intro}

\footnotetext[0]{Corresponding author: Levente Juhasz (ljuhasz@fiu.edu) \newline \newline This is a preprint version. The final article published in \textit{IEEE Access} is available at \url{https://doi.org/10.1109/ACCESS.2022.3197222}}

There is a special place on Earth at an equally interesting location. Although it has no spatial extent, it has a thriving community and digital economy: every day many people record their fitness activities, there are countless properties offered to sale and it is even the origin of malicious cyber attacks \cite{cyberattack}. Many restaurants are located there, and delivery drivers are always available to make stops at vacation rentals, there is social media activity with millions of photos uploaded, and the place even has an airline. This place is truly a  product of our digital age. It is called Null Island, and it is located at the center of the Earth. Although its reputation is growing as more and more people become aware of its existence, this paper will make a valuable contribution to raising awareness of the most interesting fact about it: that it does not exist in a way most places do. This paper will make an important contribution to the discourse of place in geographic information science (GIScience). Even though Null Island is ‘fictional’, its implications concerning geographic information are very real, and as such, Null Island and its associated issues should be discussed, in a serious and sustained manner, within the GIScience community and beyond.

\begin{figure}
  \includegraphics[width=\linewidth]{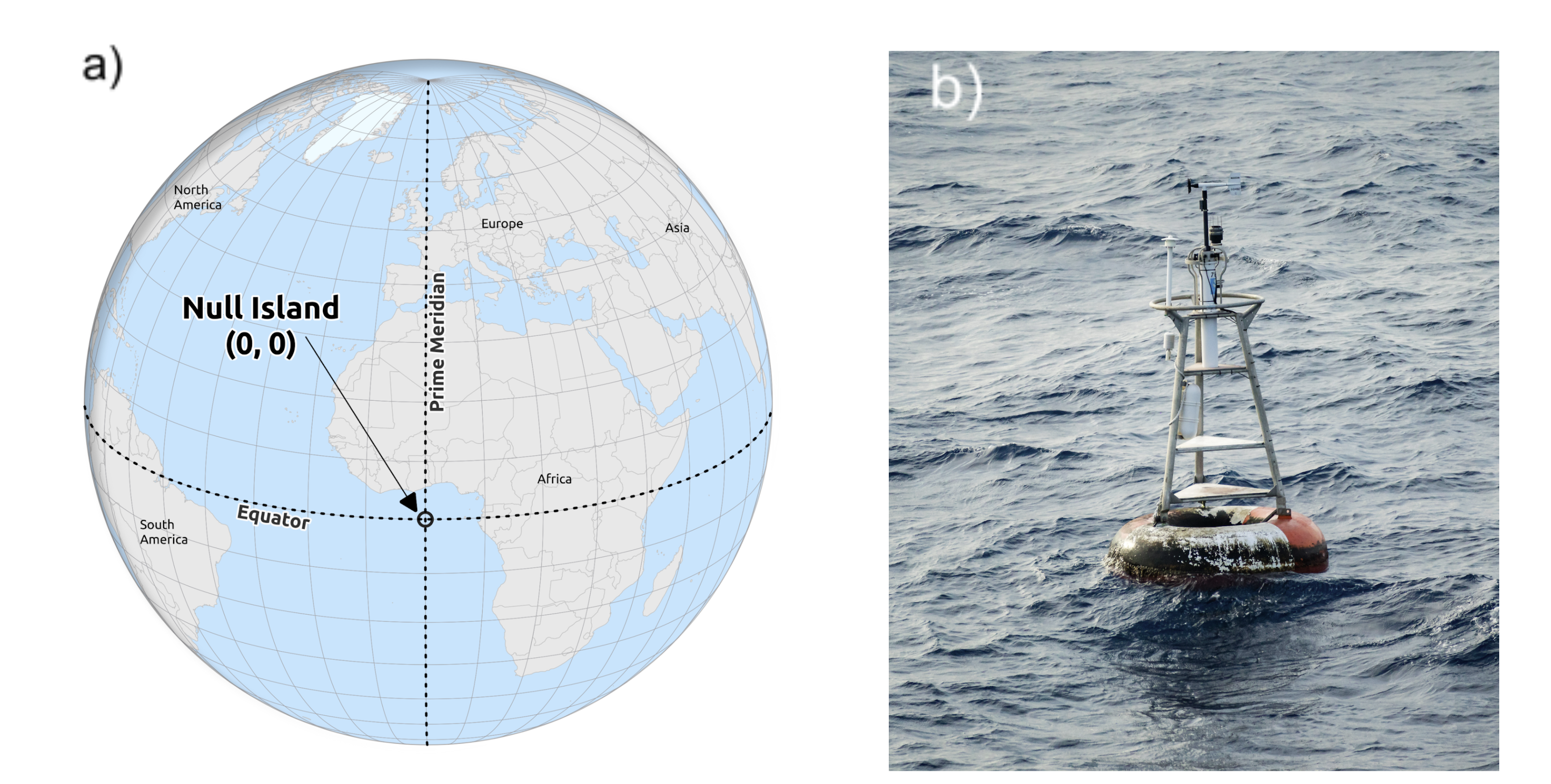}
  \caption{Null Island’s location at the intersection of the equator and prime meridian (a), and a PIRATA buoy named Soul permanently anchored to the seabed at that location (b)}
  \label{fig:where}
\end{figure}

The name Null Island is used to refer to the location on Earth where the equator intersects the prime meridian at 0° latitude and 0° longitude (0, 0) in the Gulf of Guinea off the coast of West Africa (Fig.~\ref{fig:where}a). Although a weather observation buoy part of the Prediction and Research Moored Array in the Tropical Atlantic (PIRATA) program is permanently anchored to the seabed at that location (Fig.~\ref{fig:where}b), Null Island cannot be considered a physical entity (i.e. an island). As Parker puts it: ‘\textit{outside databases, Null Island does not exist}’ \cite{parker_humble_2020}. It exists only as a placeholder for bad data in databases and digital maps. It is also regularly the topic of social media discussions (e.g. as highlighted in the title of this paper), popular media articles and blogs as well as  appearing as an artistic concept. This renders it as a real place without traditional spatial properties. Originally considered as an insider joke within the geospatial community we argue that the concept of Null Island evolved into a wider phenomenon with significant social and technological implications that reach beyond GIScience. People have always found geographic extreme points and superlatives  interesting \cite{tripoints_paper, walktheline}, and  fictious places can become real. For example, Agloe, NY was originally a 'paper town' or copyright trap in the 1930s. Following its inclusion in paper maps, the Agloe General Store opened, which was followed by a gas station and two houses, which eventually lead county administrators to consider its existence \cite{green_paper_2009, latif_caveat_2019}. However, Null Island is different in the sense that it is the product of human-computer interaction and it was discovered rather than made up. This paper will also demonstrate that far from existing as an imaginary extreme point Null Island has transformed into a fictional point that has become very real, even though it does not exist. Subsequently, we contribute to raise awareness with this structured, considered and academic treatment of Null Island as a subject. A discussion of Null Island, as is presented in this paper, is missing from the literature. We also provide guidelines on how to avoid the pitfalls and negative aspects of Null Island when accessing, visualizing, managing and conversing about geographic data and information. 

The remainder of the paper is structured as follows. The final parts of this introduction establish the importance of discussing Null Island in Section~\ref{importance} and describe the methods and materials used in the study in Section~\ref{methods}. Section~\ref{defining_null_island} defines Null Island by summarizing error sources that assign geographic data to $(0, 0)$ in Section~\ref{error_sources}, and by describing the history of Null Island in Section~\ref{history}. Section~\ref{what_is_null_island} gives an accurate representation of data that we can find on Null Island today at the time of writing, which is followed by Section~\ref{implications} that details the implications of having erroneous data on (0, 0). The findings and implications are discussed in Section~\ref{discussion}, where guidelines to avoid mistakes that incorrectly put data on Null Island are also presented (Section~\ref{guidelines}). Finally, Section~\ref{summary} concludes the work by providing a summary and directions for future research.

\subsection{The importance of discussing Null Island} \label{importance}
One of the important contributions of this article is to frame a discussion of Null Island as a fundamental and conceptual issue of geographic information. Technological advancements in the last few decades have made it very easy to create web-based maps without any or much geospatial training. This can be considered as the democratization of mapping  since creating maps is not the privilege of geospatial professionals anymore \cite{crampton_introduction_2005}. On the other hand, lack of geospatial training can easily lead to bad mapping practices resulting in faulty or even intentionally misleading maps \cite{mooney_mapping_2020} as well as errors that could otherwise be avoided by  careful mapping practice. For example, people not familiar with spatial data types, projections and other concepts are likely to ‘rediscover’ Null Island from time to time when data are incorrect and appears in the middle of the ocean. This is apparent in the amount of questions and discussions similar to \textit{'why is my data mapped off the coast of West Africa?'} in programming and GIS related Q\&A (question-answer) websites, forums and issue trackers. It is therefore important to discuss Null Island as a fundamental and conceptual issue of geographic information. We contribute detailed discussion of Null Island and its associated issues which are currently missing from the literature. We believe this omission in some ways contributes to its ‘mystical nature’. Most academic studies that mention Null Island use it in place of (0, 0) location \cite{mound_regional_2019, dijt_trajectory_2020, hill_how_2020, stein_hybrid_2020, gregoire_model_2021, hakima_low-thrust_2021, kousis_could_2022}, and only a handful studies go beyond this simplistic view to mention Null Island in relation to its original purpose: to trap geocoding errors \cite{janowicz_moon_2016, kopsachilis_geolod_2021}. Null Island’s history and purpose was also described in an architectural design article \cite{dillon_zero_2021}. Compared to these, our paper will deliver a more comprehensive treatment of what Null Island really is by describing and explaining its significance that reaches well beyond the geospatial profession. 

This article therefore contributes an innovative, informative and timely contribution to both the academic spatial community and the general public by providing a detailed account of Null Island and its associated issues. It will serve as a guide that not only describes the technical considerations of Null Island from software development and geospatial perspectives, but it will introduce the growing interest in Null Island as a social and artistic concept as well. Based on investigation, the article will also provide guidelines to avoid common pitfalls that wrongly associate geographic data with Null Island.

\subsection{Materials and methods} \label{methods} 
One of the key investigative measures employed in order to find materials about Null Island relied heavily on web searches. We utilized Google (\url{https://www.google.com}) and DuckDuckGo search engines (\url{https://duckduckgo.com}), as well as searches in various online databases and services. We strategically searched through the content of websites, blogs, databases, mailing lists, forums, news archives (e.g. NewsBank), Q\&A sites (e.g. GIS StackExchange available at \url{https://gis.stackexchange.com}) as well as social media (e.g. Twitter, Instagram) and other location-based services (LBS, e.g. Waze) to find mentions and examples of Null Island and how the term is being used on the Internet. Where available, we also utilized application programming interfaces (API) to further explore content that otherwise might not be available on the public facing web, or would be difficult to find. For example, the \texttt{geosearch} method on Wikipedia’s API is able to locate all articles on or near a specific coordinate \cite{mediawiki_geosearch_2022}. For text-based searches, by default we searched for the specific expression \textit{‘Null Island’}. This introduces a natural bias against alternative names and potentially misses relevant content. To mitigate this bias, we also included other terms referring to the same location in our searches (e.g. ‘zero latitude, zero longitude’, 0°N, 0°E, etc.). For location based queries, we searched for data on or geographically near (0, 0).

The searches mentioned above can only provide a snapshot of what is available on the internet today. However in reconstructing the story of Null Island, historical accounts must also be considered to build an accurate picture on the evolution and use of the term. For that reason, we placed a strong emphasis on investigating historical websites. More specifically, throughout our investigation we made extensive use of the Wayback Machine (WM, \url{http://web.archive.org}), which is part of the Internet Archive Project, capable of retrieving past content of websites \cite{arora_using_2016}, even if they are not available today. For example, in 2022, the domain \url{nullisland.com} is listed for sale and the original content is not available to browse anymore. However, the WM stores historical versions of the website that made our investigations possible. It is important to mention that the WM is only able to retrieve websites that were indexed at least one point in the past, and is only able to retrieve a static snapshot at the time of indexing. The WM is also used to provide permanent links for online resources cited in the article.

\section{Defining Null Island} \label{defining_null_island}
On the surface, Null Island is a technological phenomenon that is product of human-computer interaction, and a result of computerized mapping systems erroneously handling spatial information. We begin by providing a detailed description of how such errors can arise. The second half of this section reconstructs the history of Null Island by providing a timeline of important events. These events together demonstrate that the concept of Null Island evolved into a social phenomenon in addition to the technological nature of it.

\subsection{How do we ''travel'' to Null Island? - Error sources} \label{error_sources}

Geocoding is the process of turning descriptive geographic data (e.g. an address or place name) into an absolute geographic reference and practically speaking geographic coordinates \cite{goldberg_text_2007}. It is an essential function not only in GIS and scientific research but in everyday life. Given the proliferation of services for navigation, spatial search and so, on millions of geocoding requests are made daily, such as entering \textit{‘123 Main Street’} in a navigation application, or asking a personal voice assistant about the weather: \textit{‘What is the weather in Miami, FL like’}? These requests require correspondence between the descriptive information and geographic coordinates. For example, a navigation system needs to know the coordinates of \textit{'123 Main Street'} in order to find the best route to that location. Similarly, the text \textit{'Miami, FL'} needs to be converted into geographic coordinates first before a spatial query can be constructed to identify location specific weather forecasts for that location. In many situations geocoding happens in the background without users explicitly being aware of it. For example, when listing a house for sale on real estate websites, the address is converted into geographic coordinates so that the website can provide a map interface for potential buyers to browse. The geocoding process works quite well due to the emergence of high quality place and address databases and gazetteers. But what happens when geocoding fails? The address or place in question might not exist at all, or the city's name might be misspelled in the input stream. Historically, instead of returning an error, many geocoding services used default locations to revert to in case of geocoding failures. By choice, this was often the coordinates of (0, 0). Setting a default location instead of discarding data without valid or missing georeference can also happen in the codebase of applications outside geocoding services. For example, missing latitude and longitude values can be represented as empty character strings (i.e. \texttt{('', '')} array) in input streams. Empty strings might get interpreted as (0, 0) latitude - longitude coordinate pair. Then, (0, 0) in the WGS84 geographic coordinate system (the default in which web based maps get input coordinates) gets placed at the intersection of the equator and prime meridian, off the coast of West Africa.

To understand how such software-based errors can be introduced into geospatial applications, one needs to understand how data values are handled in database management systems (DBMS) and handled within data structures in  programming languages. The so-called ‘null value problem’ relates to the treatment of missing information in relational DBMS and is well known in database research. Even Codd, the inventor of the relational database model, wrote extensively on the semantics of handling missing values (see e.g. \cite{codd_extending_1979, codd_missing_1986}). A potential problem with ‘null values’ is that their presence introduces three-valued logic, where not only \texttt{true} and \texttt{false} boolean values are possible. There exists a third unknown or undetermined value. This is problematic since it introduces a new degree of uncertainty in attribute values. Furthermore, the treatment and evaluation of ‘null values’ depend on the DBMS \cite{thalheim_null_2011} and there are also inconsistencies within the Structured Query Language (SQL) standard \cite{van_der_meyden_logical_1998}. Since whether a \texttt{NULL} value is interpreted as zero, \texttt{FALSE} or \texttt{NULL} depends on many factors, the ‘null value problem’ is considered a programmer or software developer's nightmare \cite{puasuareanu2009survey}. Outside databases, JavaScript, the language of most web mapping platforms, works with a variety of data types (i.e. \texttt{numeric}, \texttt{boolean}, \texttt{string}, etc.), and also allows the programmer to convert one to another through methods called type conversion and type coercion (implicit conversion). For example, both \texttt{parseFloat()} and \texttt{Number()} built-in functions are designed to convert data to the \texttt{Number} data type. Both functions return the number 0.0 when converting the character \texttt{'0'} to a number (Table~\ref{tab:javascript}, line 1). In other cases, however, different behavior can be observed, for example when passing an empty character string (\texttt{‘’}) to these functions (Table~\ref{tab:javascript} line 2). The empty character string is converted into a \textit{NaN} (Not a Number, a special data type) by the \texttt{parseFloat()} function, and into the 0.0 number by the \texttt{Number()} function. This case is quite common if the input data were originally stored in a comma separated text file (CSV) and had missing geospatial coordinates. Another example can be converting a character string that starts with the \texttt{‘0’} character (Table~\ref{tab:javascript}, line 3), which is converted into 0.0 and \textit{NaN} by \texttt{parseFloat()} and \texttt{Number()} respectively. This example can happen due to misspelling values when recording the original data. The behavior of these two functions can be confusing, especially if the programmer is not familiar with JavaScript, or is new to programming in general. In addition, more confusion can be introduced by the process of type coercion, which is a special type of type conversion that happens implicitly during operations performed on two different data types - sometimes without the programmer realizing this behavior is happening.

\begin{table}
\caption{Examples of explicitly converting data (type conversion) to floating point numbers in JavaScript with the \texttt{parseFloat()} and \texttt{Number()} built-in functions}
\setlength{\tabcolsep}{3pt}
\begin{tabular}{|l|cc|c|} 
\hline
   Input value & \multicolumn{2}{l|}{Converted value} \\ 
   
    & \texttt{parseFloat(}\textit{Input}\texttt{)} & \texttt{Number(}\textit{Input}\texttt{)}   \\ 
\hline
  \texttt{'0'} & \texttt{0.0} & \texttt{0.0}  \\
 \texttt{''} & \textit{NaN} & \texttt{0.0}   \\
\texttt{'0asd'} & \texttt{0.0} & \textit{NaN}  \\
\hline
\end{tabular}
\label{tab:javascript}
\end{table}

Geocoding and programming issues are not the only way for data points to get placed on Null Island. Table~\ref{tab:sources}. summarizes ways in which data can end up  positioned on Null Island. Apart from intentional use of Null Island either as a joke or as a container for data, most issues presented in the table are problematic from a geospatial perspective because of the geographic offset between (0, 0) and the real-world location the data point intends to represent. The most common implications of bad data will be discussed in Section~\ref{implications} in more detail.

\begin{table*}
\caption{Error sources that place data on Null Island}
\setlength{\tabcolsep}{3pt}
\begin{tabular}{|l | p{3.5in} | p{1.9in} | }  
\hline
 \textbf{Issue source} & \textbf{Description} & \textbf{Example} \\ 
 \hline
 Default location & Instead of not returning data in case of failures, returning (0, 0) as the default location. This behavior can be a result of geocoding, or the programmed behavior of applications & Multiple addresses georeferenced to Null Island in Fig.~\ref{fig:examples}e.    \\ 
 \hline
 Programming issues & Unintentional programming errors, such as data type conversions or error handling that lead to (0, 0). This can be a result of an inexperienced programmer, lack of testing, issues with the input data (e.g. missing values), or the combination of these. & Newline characters preceding the final \texttt{</coordinates>} tag in KML files rendered the last point of coordinate arrays in Null Island\cite{google_code_google_2011}  \\ 
 \hline
 Hardware/software issues & Geolocation methods (e.g. GPS/GNSS receivers, wireless geolocation, cell triangulation) failing to obtain fix coordinates and report (0, 0) instead & A digital image with the \texttt{GPS Position} EXIF tag: \texttt{ 0° 0' 0.00" N, 0° 0' 0.00" E}   \\ 
 \hline
 Projection issues & Latitude - longitude coordinate pairs (in degrees) interpreted as projected coordinates (e.g. in EPSG:3857) & 25.7$^{\circ}$N, 80.2$^{\circ}$W (Miami, FL) would be rendered 25.7m North and 80.2m West from Null Island in the WGS84/Pseudo-Mercator (EPSG:3857) projection \\ 
 \hline
 Intentional use & Data is deliberately associated with the coordinate (0, 0), due to, for example users' desire for increase privacy (e.g. location spoofing); explicitly referencing Null Island as a place or joke & An Instagram post about ‘Air Null’ a fictional, humorous airline (Fig.~\ref{fig:airnull})   \\ 
 \hline
Container for data & Using Null Island's concept as a deliberate container for data with no or uncertain geographies & See Section~\ref{container} \\
\hline
\end{tabular}
\label{tab:sources}
\end{table*}

\subsection{A brief history of Null Island} \label{history}
\begin{figure*}
  \includegraphics[width=\textwidth]{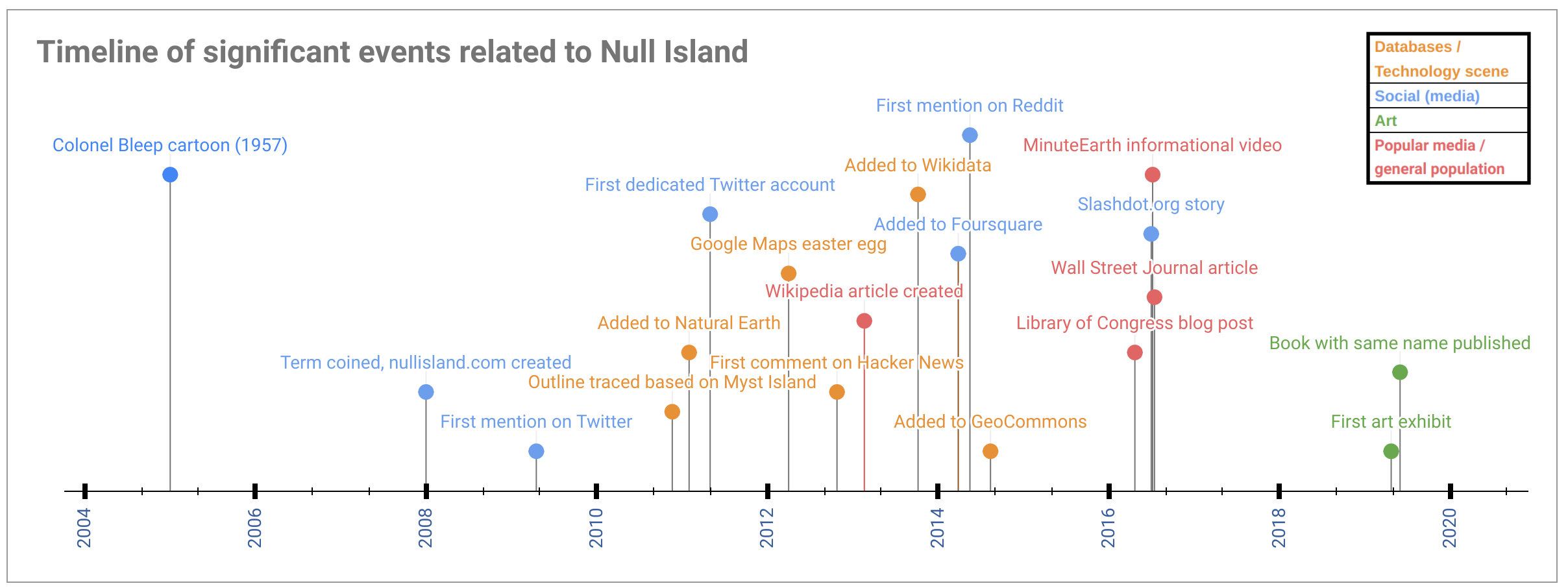}
  \caption{Timeline of important events related to Null Island}
  \label{fig:timeline}
\end{figure*}

In order to describe the history of Null Island we have constructed a visual timeline of important events associated with Null Island as shown in Fig.~\ref{fig:timeline}. The figure plots different types of events (i.e. related to technology and databases, and social aspects including social media, general population) with different colors. The timeline suggests that the term `Null Island' was indeed originally within the geospatial community  as an insider joke since all events that we were able to identify within four years after the first mention are related to individuals identifying themselves as geospatial professionals. Upon closer examination of these events, it appears that Null Island as a concept managed to break away from the geospatial community and establish itself within the context of larger groups, as indicated by multiple events in the database/technology and general population categories. We identified four evolutionary phases in the timeline of Null Island based on our estimation of the interest the term received from various communities. These timelines are fuzzy but we believe they provide good anecdotal guidance to the phases of evolution of Null Island. The four phases are as follows:

\begin{enumerate}
  \item \textbf{Phase 1}: Known within the geospatial community: ``Like no place on Earth'' - An insider joke (2008 - Early 2010s)
  \item \textbf{Phase 2}: Increased popularity and acceptance within the larger technology community (Early 2010s to mid-2010s)
  \item \textbf{Phase 3}: Null Island is now successfully reaching out into the general population (Mid-2010s to late 2010s)
  \item \textbf{Phase 4}: Moving beyond technology only - the concept of Null Island is transformed into an artistic concept (Late 2010s to present)
\end{enumerate}

References to some of the events omitted from the following subsections and Fig.~\ref{fig:timeline} are provided as an appendix in Table~\ref{tab:first_events}.

\subsubsection{Phase 1: ''Like no place on Earth'' - An insider joke} \label{joke}

\begin{figure*}
  \includegraphics[width=\textwidth]{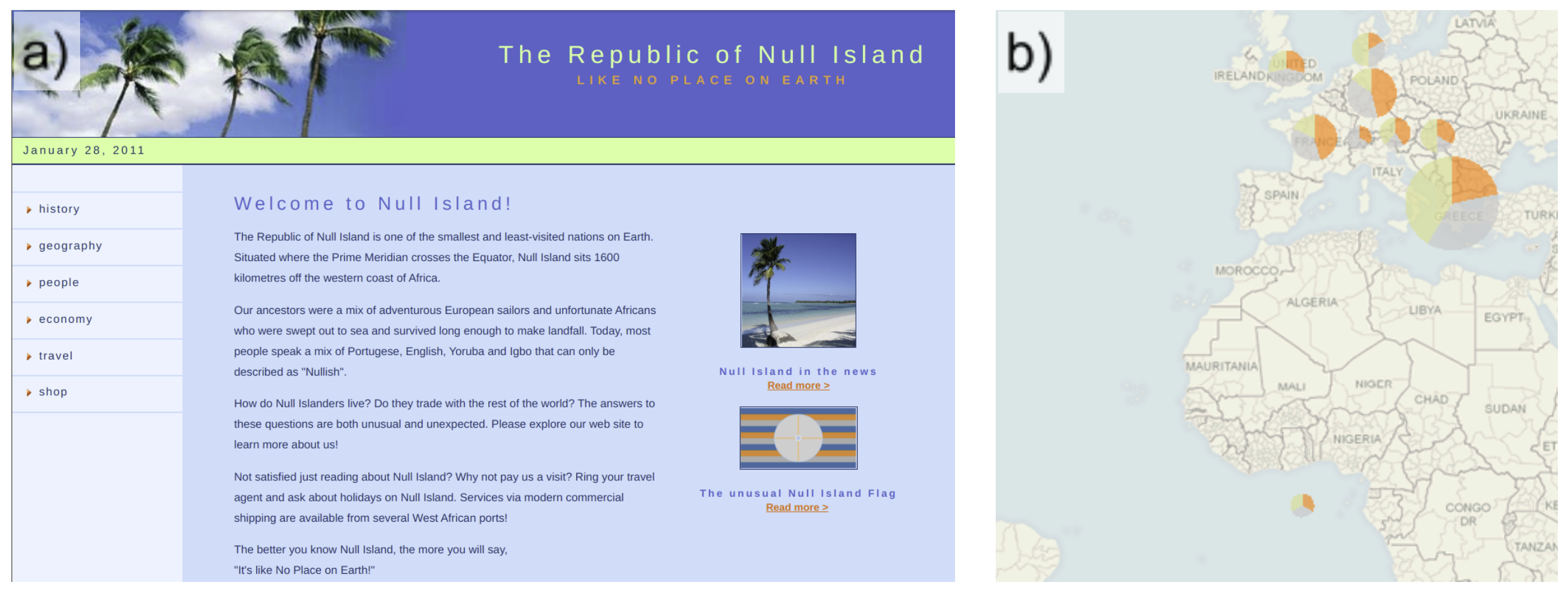}
  \caption{First examples of digital content dedicated to Null Island; a website dedicated to the fictional ‘Republic of Null Island’ (a), and a map showing olympic medal counts of the 1896 summer olympics erroneously rendering data on Null Island (b)}
  \label{fig:website}
\end{figure*}

Most sources agree that the term `Null Island' was coined in 2008 by Steve Pellegrin, a data analyst who at that time worked at Tableau, a data visualization software company. His intention with the term was to \textit{'describe data goofs'}. In the same year, a website dedicated to the fictional Republic of Null Island was created (Fig.~\ref{fig:website}a). Historical DNS (domain name system) records show that the 2008 version of the domain name was owned by Steve Pellegrin (Fig.~\ref{fig:whois}), which confirms that Null Island was originally coined by him. The website created   a backstory detailing the island's history, geography and economy \cite{pellegrin_republic_2011}. According to Pellegrin, the island is inhabited by $4,000$ people, it has the highest number of Segway scooters per-capita in the world, and most of the working age population works in the software development industry. The website clearly establishes Null Island as a joke, which is further reinforced in a 'blog-like' section of the site where ‘fun-facts’ are reported, such as erroneous maps showing data mapped to Null Island (Fig.~\ref{fig:website}b). The site also featured a webshop, selling Null Island branded merchandise, such as coffee mugs, t-shirts and hats. The first mention of Null Island on Twitter appears to be from 2009, as a response to a user asking about being wrongly geolocated \textit{'off the coast of Africa'}. The response tweet by Christopher Currie (senior software developer at Tableau) reads as \textit{'We call that spot Null Island'}, which implies that by that time the term was established at Tableau to refer to (0, 0). The first Twitter account dedicated to Null Island (\texttt{@nullisland}) was created in May 2011 in the same platform, with multiple other ones following (e.g. \texttt{@nullislandgang}, \texttt{@nullislandbouy}, and \texttt{@MaptimeNull}).

At this stage, with the exception of the \url{nullisland.com} website, Null Island appeared as a dimensionless object at the center of the world. An important step in establishing Null Island as a more widely accepted fictional place was providing it with more elaborate spatial properties in the form of an outline. The first version was created in 2010 by GeoIQ and Stamen Design, when they included it in their newly designed basemap style called Acetate \cite{migurski_acetate_2010}. The outline was based on the major island from the video game Myst, which at its peak popularity in the 1990s was the best selling video game in history. Perhaps one of the most significant events in the real history of Null Island happened in early 2011, when it was added to version 1.3 of the Natural Earth (NE) database \cite{natural_earth_natural_2011} as a troubleshooting country. The intended purpose of adding a 1 m$^{2}$ polygon centered at (0, 0) was to flag geocoding failures, which at the time were routed to (0, 0) by most mapping services. The feature has a scale rank (a measure roughly corresponding to web map zoom levels) of 100, indicating that it should never be rendered in maps, but should only be used during analysis to keep errant points off maps \cite{natural_earth_natural_2011}. The significance of Null Island's inclusion in NE is that the dataset is in the public domain making it part of one of the most popular sources for geographic data. This large scale dataset has been downloaded close to seven million times as of 2022 February. We argue that this has provided Null Island with unprecedented exposure and opportunities to be discovered and explored with many applications that require geographic data incorporating NE.

\subsubsection{Phase 2: Acceptance within the larger technology community} \label{tech}

\begin{figure*}
  \includegraphics[width=\textwidth]{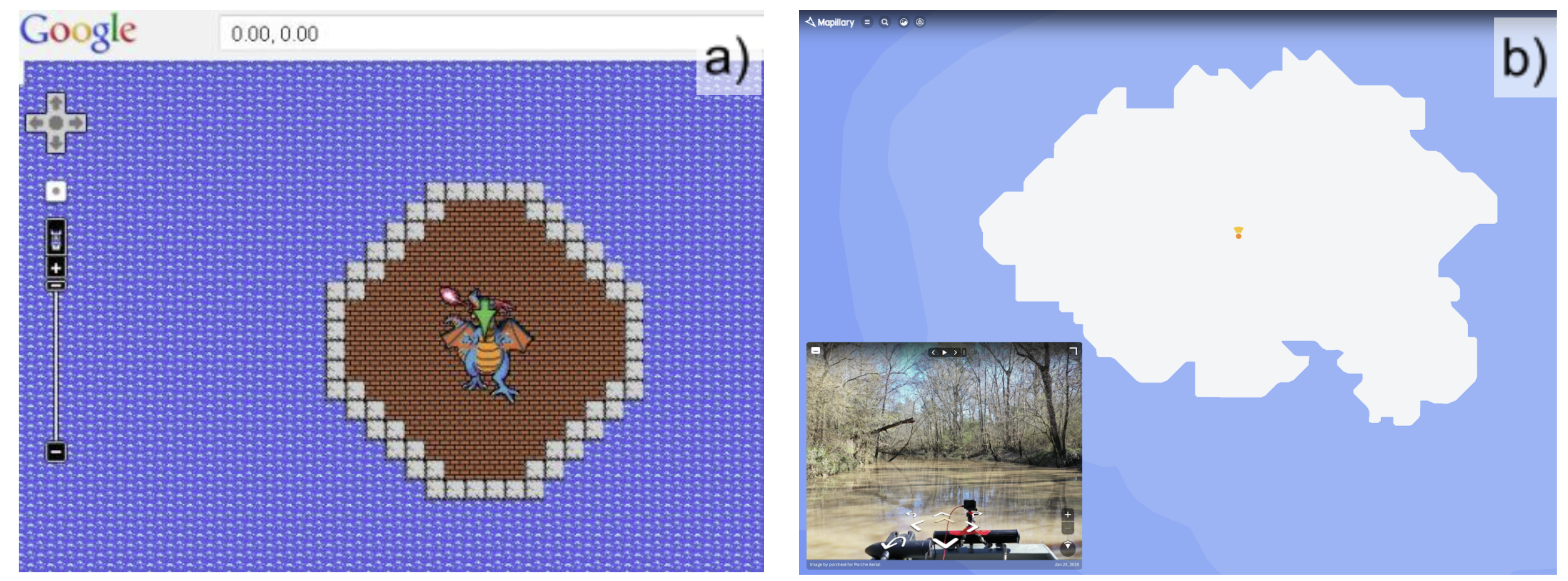}
  \caption{Null Island and its more elaborate spatial properties by the technology community: Google Maps Easter Egg featuring Null Island in 2012 (a), and Null Island’s outline based on Myst Island as appears on Facebook’s map products in 2022 (b)}
  \label{fig:outline}
\end{figure*}

Events in  this phase allowed Null Island to reach an audience beyond the geospatial community and be known within the larger technology landscape mainly consisting  of developers, data scientists and open data advocates. Adding the representation of Null Island to databases not only continued after NE but also broadened in scope. In addition to appearing in other geospatial datasets, like Geocommons in 2014, Null Island was included in general collections, such as Wikipedia in 2013 and Wikidata in 2014. Even though the nature of Null Island is playful, it started to gain significance as a more serious concept. For example, \textit{Who's on First}, a gazetteer project that aims to represent all places in the world assigned the distinguished permanent ID of 1 to Null Island (\url{https://spelunker.whosonfirst.org/id/1/}), and occasionally, the very first node in the OpenStreetMap (OSM) database also gets moved to (0, 0) (See \url{https://openstreetmap.org/node/1/history}, versions \#15 and \#17). Similarly to Twitter, other social media outlets as well as social news aggregator websites that were popular among programmers also mentioned Null Island for the first time between 2012 and 2016 (Fig.~\ref{fig:timeline} and Table~\ref{tab:first_events}). In 2012, Google released a so-called Easter egg version \cite{noauthor_easter_2022} of Google Maps closely resembling 8-bit video games which allowed users to explore the world in 8-bit \cite{nomura_begin_2012}. This special version of Google Maps included an imaginary island at (0, 0), borrowed from the computer game Dragon Quest (Fig.~\ref{fig:outline}a). The original 2010 outline of Null Island by GeoIQ and Stamen has also been reworked by GNIP (a technology company later acquired by Twitter) in 2013 as a SVG file that allowed wider adaptation of the outline using standard graphic software\footnote{ The repository that contains the SVG image is available at \url{https://github.com/gnip/null-island}} and subsequently included on more merchandise, such as T-shirts and stickers. Furthermore, the addition of the outline to GeoCommons (now based on Myst Island's shape) was built into numerous web basemaps. For example, Facebook's map products still render the outline of Null Island in 2022, as seen for example on Mapillary (now owned by Meta, the parent company of Facebook) in Fig.~\ref{fig:outline}b.

\subsubsection{Phase 3: Successfully reaching the general population} \label{general_pop}

\begin{figure}
  \includegraphics[width=3.5in]{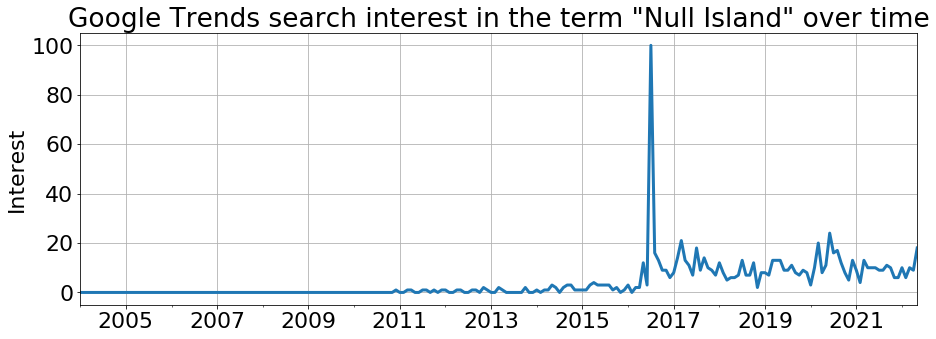}
  \caption{Search interest of the term ‘Null Island’ on Google Trends shows peak popularity in 2016 and sustained interest since then}
  \label{fig:trend}
\end{figure}

We observe that the idea of Null Island was able to reach beyond the geospatial community and spark the interest of larger technologically oriented audiences from its role in many conversations across the Internet.  In order to judge the significance of Null Island during this phase, a natural next question is  whether it had reached a general familiarity within the general population that did not necessarily have background and interest in GIScience or computer science-related disciplines. To assess this, we utilized Google Trends and extracted the search interest in the term 'Null Island' over time, which shows the extent of web searches that were conducted on Google\footnote{ The service is accessible at \url{https://trends.google.com}. We used the following query parameters to extract worldwide search interest: \texttt{date=all\&q="Null Island"}. The query can be reproduced at \url{https://trends.google.com/trends/explore?date=all&q="Null Island"}}. Google does not show the absolute number of web searches, but rather provides an index of interest from 0 to 100, calculated within a given period, where 100 is the peak popularity of the search term. An interest of 50 means that the search was half as popular for that specific time. The value of 0 means not enough data was available to calculate the index \cite{ballatore_tracing_2020}. Google also lists other searches made by users who were interested in Null Island. The top 5 related queries were ‘null island google maps’, ‘null island buoy’, ‘null island flag’, ‘0 0 coordinates’ and ‘null island t shirt’ in order of decreasing popularity. This list of related queries suggests that search interest shown in Fig.~\ref{fig:trend} represents interest in Null Island, and not the result of random noise from web searches.

The previous two phases show  how Null Island gained popularity within the geospatial and wider technology communities, however, Fig.~\ref{fig:trend} suggests that this did not infiltrate the general population until 2016. There is evidence of slight, sporadic interest before 2016 for example a mention in The Sunday Herald in Glasgow, Scotland in 2014 \cite{jameison_into_2014}. However the majority of search interest values remain 0 or very low. The peak popularity was reached in 2016. By this time, the original 2013 Wikipedia article was translated into major languages such as German, Spanish, French, Italian and Russian. By 2022, this list has grown to 17 languages including Chinese, Japanese, Arabic and Portuguese among others. The peak popularity in 2016 can be traced back to popular media outlets reporting stories on Null Island. We found evidence of at least three independent major contributions that were picked up and shared by many other media and social media sources numerous times. First, the Library of Congress  released a blog post in April 2016 titled 'The Geographical Oddity of Null Island’ \cite{st_onge_geographical_2016}. This was followed by an informational video released by MinuteEarth on Youtube titled 'Null Island: The Busiest Place That Doesn't Exist' \cite{minuteearth_null_2016}. This video has accumulated over 2.2 million views as of time of writing in 2022. To further add to the list of popular outlets reporting on Null Island, The Wall Street Journal wrote ``If you Can't Follow Directions, You'll End Up on Null Island'' which appeared both in print and online \cite{lee_hotz_if_2016}. These media mentions popularized the idea of Null Island to a general audience, which is apparent in a more sustained web search interest seen in Fig.~\ref{fig:trend} after 2016.

\subsubsection{Phase 4: Discovery by the arts - now more than just technology} \label{arts}

Long before the geospatial revolution, the first color cartoon ever made for television, Colonel Bleep was created in 1957 \cite{beck_colonel_2018}. The show featured a prominent location as Colonel Bleep’s headquarters called Zero Zero Island, which was located where the equator meets the prime meridian (Fig.~\ref{fig:visual_arts}a). Although this fictitious land mass is not identical to Null Island, this shows that  interest in the `center of the Earth' by the visual arts is not new and that (0, 0) was considered a special location \cite{timothy1998collecting}.

Recently, other forms of arts have also discovered Null Island as an artistic concept. Artists have started using it as a metaphor and a myth where all lost objects (i.e. without locations or coordinates) are collected. The interest seems to stem from the paradox that Null Island is a tangible, known location that exists in place databases and in common knowledge, yet, it is non-existent technically~\cite{RoseImagined10.1080/09502389700490011}. Therefore, the idea of blending imaginary and real geographies seems to get more popular since 2019 when the first art exhibition dedicated to Null Island called 'NULL ISLAND: Exploring the Busiest Place on Earth that Doesn't Exist' was held by ADS4 at the School of Architecture at the Royal College of Art, London. In another exhibition, Mapping the Cartographic in 2020, Deborah Mora creates a video installation called ‘0° N, 0° E’ designed to retrace the origins of Null Island and collect digital materials from biologist, geologists, such as satellite images, 3D models, photos, etc. that represent nature to create her video\footnote{ The archived version is available at \url{http://web.archive.org/web/20210123224002/https://nextmuseum.io/en/submissions/0-n-0-e/}}. Mora notes that 'Null Island becomes a timeless, liminal place where all these objects try to survive virtually, beyond material deterioration’. Another artistic representation of Null Island by Letícia Ramos is shown in Fig.~\ref{fig:visual_arts}b. Null Island also lends its name to a book \cite{moreno_null_2019} and a collection of poems called Null Landing \cite{hines_null_2022}. These artistic representations of Null Island open up a larger discussion about fiction, geography and literary nonsense that is beyond the scope of this paper.

\section{What is Null Island today?} \label{what_is_null_island}

The previous sections have described the technical details of how data can become associated with Null Island. We have also carefully demonstrated the growing interest and  awareness about Null Island over the last number of years. This section illustrates why Null Island is often seen as the busiest or most popular place on Earth.

\subsection{A placeholder for ''bad'' data} \label{bad_data}

Nicknames have been assigned to Null Island due to the amount of different data that accidentally end up on Null Island. These data span across databases and datasets and affects some of the world's most popular location based services with hundreds of millions of users. This indicates that ``bad data'' on Null Island is not an isolated case. Naturally, a lot of technological advancements have taken place since Null Island was first coined in 2008 which influences the types of examples we encounter. For example, newer, improved versions of mapping and geocoding software are more likely to have improved error handling and more reliable results therefore, in our opinion, the likelihood of wrongly locating data on Null Island has probably decreased since 2008.

\begin{figure*}
  \includegraphics[width=\textwidth]{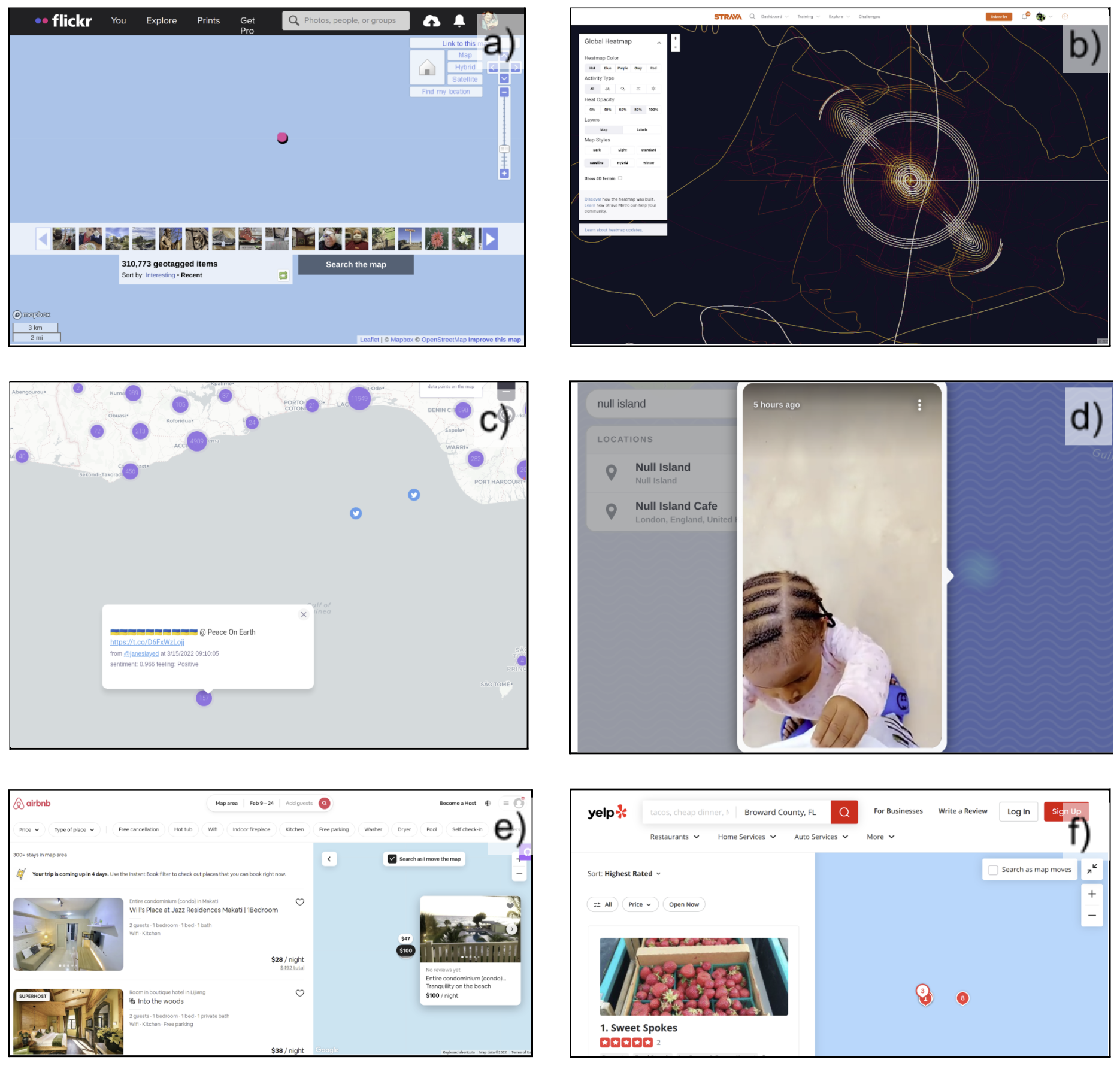}
  \caption{Recent examples of data wrongly associated with Null Island are present across different geospatial apps and services: (a) Flickr geotagged photos, (b) Strava fitness activities, (c) geolocated Tweets, (d) Snaps on Snapchat as well as the location of Null Island is searchable, (e) AirBnB vacation rentals and (f) Yelp venues (restaurants and other POIs)}
  \label{fig:examples}
\end{figure*}

Most LBS mainly rely on smartphone geolocation. Smartphone geolocation does not only use satellite geopositioning. Other methods using wireless and cell phone triangulation are also used to determine the geographic location of a device \cite{merry_smartphone_2019}. Even though these methods are designed to be redundant, poor or no position can still affect user generated data. This leads to contents such as photographs and other geotagged content being associated with (0, 0). Such problems span a wide variety of applications, such as geotagged photo services, fitness activity trackers, and more traditional social media outlets. Fig.~\ref{fig:examples}a shows over $300,000$ geotagged photos on Flickr geolocated on Null Island. Unlike Flickr, that accepts positional data as is without checking, other services may post-process photographs, such as Mapillary \cite{juhasz_user_2016} where 3D scene reconstruction from overlapping images is used for quality checking. Despite this the issue of locating photographs on Null Island remains as illustrated in Fig.~\ref{fig:outline}b. Fitness tracker software are another type of application relying on geolocation from smartphones and other devices, such as GPS-enabled smart watches. Strava, a leading global company in this area, publishes a global heatmap that aggregates user activities to show areas that are favored by the its fitness or user community \cite{kulyk_determining_2018}. Fig.~\ref{fig:examples}b features a screenshot from this global heat map centered at Null Island, that shows that many fitness activities are being uploaded with coordinates on or near (0, 0). Social media data that are commonly used in GIScience are also affected, such as Twitter (Fig.~\ref{fig:examples}c) and Snap Map, Snapchat’s map interface \cite{juhasz_analyzing_2018}. Fig.~\ref{fig:examples}d shows that Snapchat not only displays posts on Null Island, but the location is also searchable by a built-in geocoding service that lets users zoom to a location of their choosing. AirBnB, the most popular peer-to-peer property rental service, currently lists over 300 vacation rentals in the Gulf of Guinea (Fig.~\ref{fig:examples}e). Yelp allows users to review restaurants and other establishments while also providing a POI (point of interest) database. Previously, it was found that business oriented services similar to Yelp provided reliable POI positions on the local scale \cite{hochmair_data_2018}, but as Fig.~\ref{fig:examples}f shows these services are not immune to geocoding errors and one can find several restaurants appearing on and near Null Island.

\begin{figure*}
  \includegraphics[width=\textwidth]{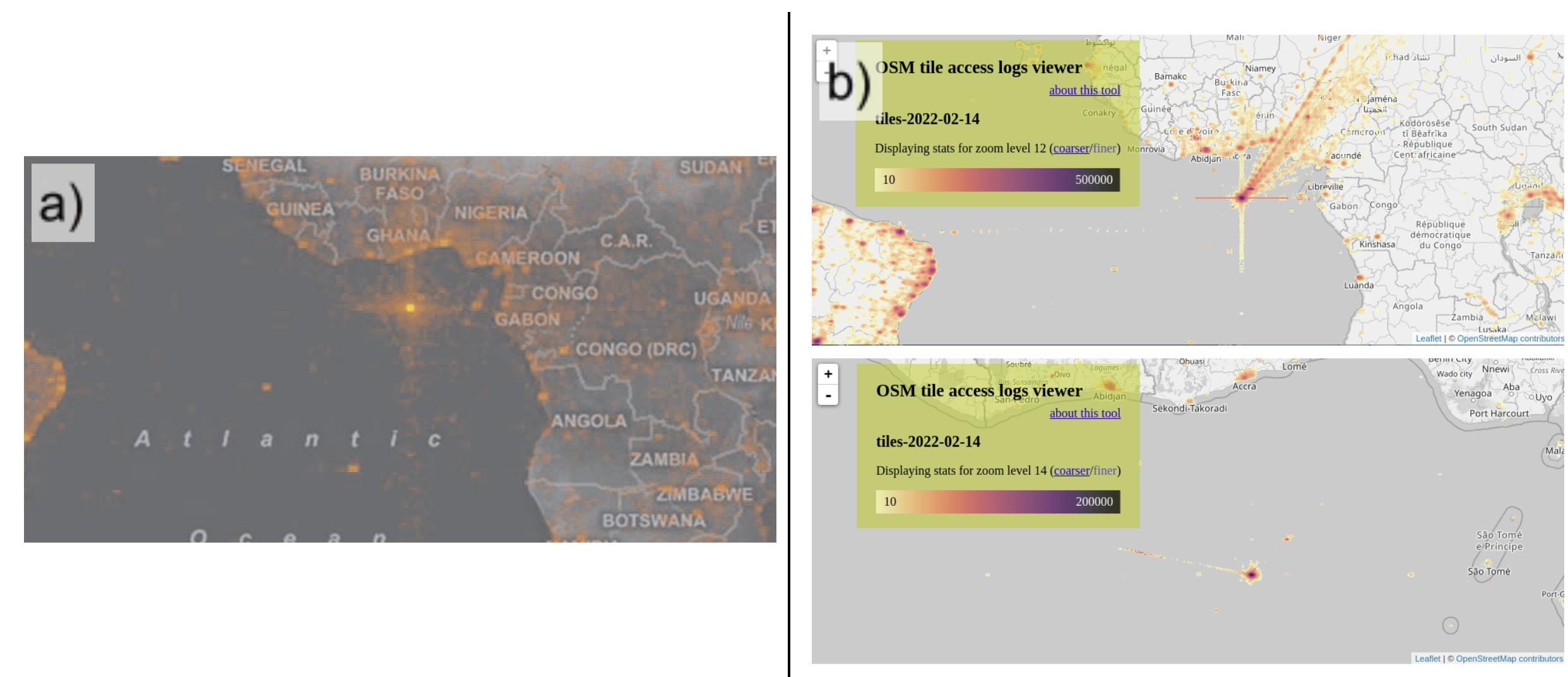}
  \caption{Visualizing popular areas that web map clients load (i.e. users see) reveal increased usage near Null Island as seen in Microsoft’s Hotmap (a); and OSM tile access visualized on level 12 (upper) and level 14 (bottom) (b)}
  \label{fig:tiles}
\end{figure*}

It is not only spatial data that contribute to this popularity of Null Island. Interactive web maps play an important role in make this location visible. Microsoft’s Hotmap (used to visualize map viewing patterns) was the first such web-based map that popularized the study of how people interact with online maps. Plotting the location of map requests, that is, map tiles that were loaded by web mapping software and viewed by users (Fig.~\ref{fig:tiles}a). This actually helped reveal a software bug that accidentally sent users to (0, 0) \cite{fisher_hotmap_2007}. Badly configured maps still exist today. We confirmed this by visualizing the tile access logs of OSM\footnote{ Tile access logs can be downloaded from \url{https://planet.osm.org/tile_logs/} and visualized using osm-tile-access-log-viewer available at 
\url{https://github.com/tyrasd/osm-tile-access-log-viewer}}. Fig.~\ref{fig:tiles}b shows the map tile usage statistics of the default tileset displayed on \url{openstreetmap.org}, which is also loaded by many other applications. Maps in Fig.~\ref{fig:tiles}b highlight areas with darker colors that are viewed more often. The upper size of the figure visualizes map tiles on zoom level 12 (ideal display for city/town overview maps), whereas the lower portion of figure uses zoom level 14 (more detailed view where individual suburbs and roads are also visible and distinguishable). Null Island is prominently displayed across all zoom levels (others not shown in Fig.~\ref{fig:tiles}b) which suggests that this is not an isolated issue. A wide array of web maps using the default OSM tiles suffer from some sort of misconfiguration that force them to load areas on or near Null Island where otherwise there would have been no data to actually see. 

\subsection{A container of data with missing or uncertain geographies} \label{container}

\begin{figure*}
  \includegraphics[width=\textwidth]{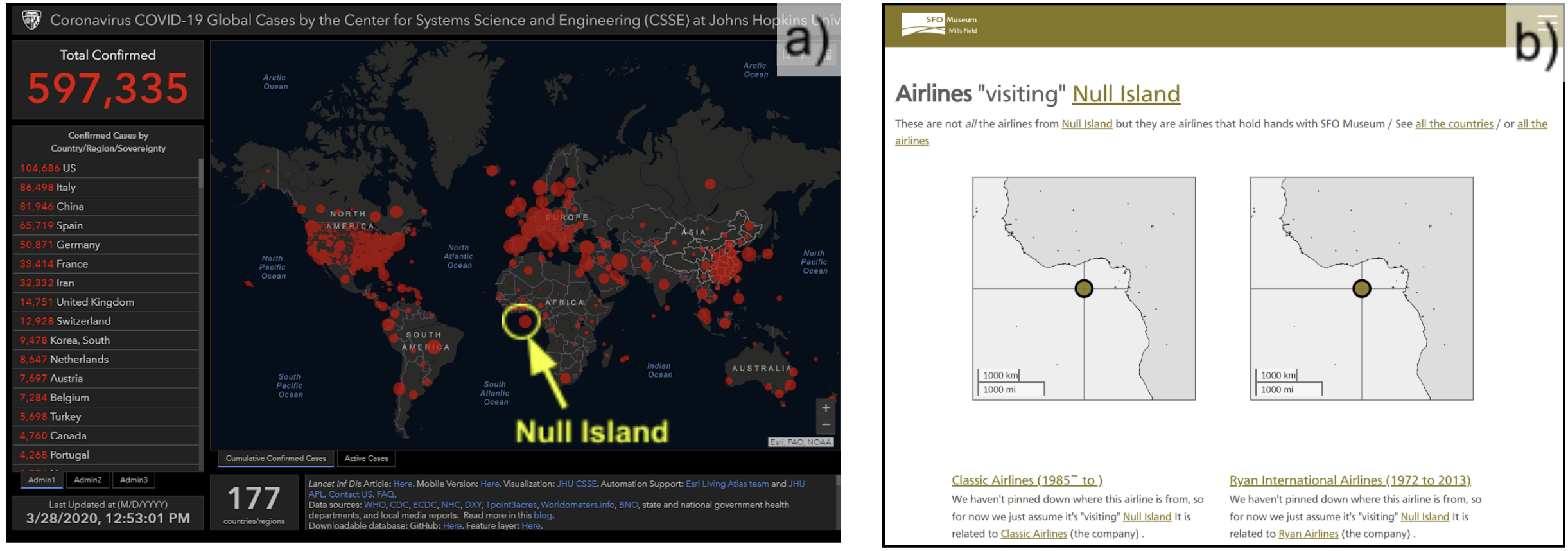}
  \caption{Null Island as a liminal place containing data with uncertain geographies in a COVID-19 dashboard developed by Johns Hopkins University (a) and in the catalog of SFO Museum (b)}
  \label{fig:container}
\end{figure*}

Following its inclusion in the NE dataset, Null Island was widely considered as a ‘placeholder for bad data’. However, the container or placeholder concept can be generalized to purposefully include data with no or uncertain geographies. An example of this is the widely popular COVID-19 dashboard developed by Johns Hopkins University \cite{dong_interactive_2020} that plotted confirmed cases on a world map during the Coronavirus disease 2019 (COVID-19) pandemic. An early version of this dashboard intentionally mapped cases with unassigned locations on Null Island (Fig.~\ref{fig:container}a). However, this behavior was changed later when the creators realized that mapping uncertain geographies without representing this uncertainty is questionable cartographic practice \cite{mooney_mapping_2020}. This was confirmed in an interview with the creators of the dashboard: \textit{‘I thought it was a great place to put everything that doesn't have a specific location yet. But that upset a lot of people, so that's gone’} \cite{keiser_every_2020}.

In the case of intentionally using Null Island as a container for data with no or uncertain geographies, Null Island can be considered as a liminal place \cite{turner_dramas_2018,Liminaldoi:10.1080/14616688.2019.1648544} or uncertain non-place \cite{auge_non-places_1995-1}.  These describe a state of uncertainty and in-betweenness.  Phyisical places are liminal when they are transitory in nature and their purpose is to connect other places. They are not destinations as people are not supposed to stay there for long. Examples include bus stations, streets or airports \cite{HUANG20181}. However, these spaces are not necessarily physical. Null Island's liminality comes from its essence in connecting the imaginary or uncertain with the real world. Reference \cite{cope_whos_2018} uses this concept as a device to signal (or contain) places with uncertain geographies in the catalog of the San Francisco International Airport (SFO) museum. Fig.~\ref{fig:container}b shows this functionality implemented in the SFO Museum’s digital collection. The figure highlights two now defunct airlines associated with SFO (e.g. once operating a route to or from the airport), with unknown countries of origin. Therefore, the airlines are cataloged to ``visit'' Null Island until the origin country can be identified and coded in the database. The catalog contains several other airports, airlines and companies ``visiting'' Null Island (\url{https://millsfield.sfomuseum.org/countries/1)}\footnote{ Note that based on the URL, the country ID of Null Island is 1, which indicates that the catalog utilizes Who’s on First (see Section~\ref{tech})}. 

\subsection{Null Island equivalents in other coordinate systems} \label{crs}

As we have discussed so far, Null Island is a place located at the origin of the WGS84 geographic coordinate system, in the Gulf of Guinea off the coast of West Africa. The geographic significance of this location is different from that of other geographically famous locations such as superlatives~\cite{Varnajot} and extreme points \cite{Looytynoja_2008} (southernmost, tallest points, etc.) In this way the location of Null Island is arbitrary and is dependent upon the WGS84 datum and geographic coordinate system that arbitrarily chose its prime meridian. Contrary to popular belief, the meridian crossing the Royal Observatory in Greenwich is located 102 m west from the zero meridian used by modern satellite navigation receivers, that use geocentric reference frames, and their realizations of the WGS84 and the International Terrestrial Reference Frame (ITRF) \cite{malys_why_2015}. Subsequently Null Island’s location is also laterally offset from the Greenwich meridian historically referred to as ``the prime meridian of the World''. Theoretically, there can be as many Null Islands as coordinate systems. To illustrate this, \cite{field_nill_2014} created a web map showing alternative Null Islands by plotting the origins of all geographic and projected coordinate systems supported by ESRI (Fig.~\ref{fig:nillponts}). However, none of these other locations gained significance nor are used in contexts similar to that of the ``original'' Null Island. This is directly related to two factors. Firstly because WGS84 is the de facto standard input coordinate system in JavaScript-based web mapping software and secondly that the WGS84 geodetic datum and geographic coordinates are used by consumer-grade global navigation satellite system (GNSS) receivers such as GNSS chips found in modern smartphones\footnote{ Note that different GNSS use different geodetic datums. GPS satellites are tracked in WGS84, but GLONASS uses PZ 90. However, for the end user, especially in consumer grade receivers such as smartphones, coordinates are reported in WGS84. See e.g. \cite{henning_user_2014}}.

Even though most web mapping platforms rely on the Web Mercator projection for rendering purposes and on the WGS84 geographic coordinates for data input, there are other alternatives. For example, D3.js (data-driven documents, \url{https://d3js.org/}), a popular JavaScript data visualization library for the web generates SVG graphics from data as opposed to rendering map tiles. The origin of the viewport coordinate system of SVG images is located at the top-left corner \cite{dahlstrom_scalable_2011}. If D3.js is used incorrectly (e.g. by an inexperienced developer not familiar with cartographic projections), converting between spatial and pixel coordinates can result in \textit{NaN} values which are placed at the coordinate system origin at the top-left corner by the SVG renderer. This is the same issue causing web maps to render bad data at Null Island (see  Section~\ref{error_sources}). This issue is present in practical programming related Q\&A (see e.g. \cite{stackoverflow_stackoverflow_2018}) which suggests that the top-left corner as a location is fundamentally similar to Null Island as it is caused by technical issues in mapping or visualization systems. However, unlike 0° latitude and 0° longitude, the (0, 0) in the SVG coordinate system (even if the image represents a map) does not correspond to one specific location on Earth. The real world location of the SVG’s (0, 0) depends on the current map image. 

\section{Implications of Null Island} \label{implications}

At this point we have considered the definition, history, evolution and a consideration of what Null Island represents today. In the following sections we consider what are the implications of Null Island in terms of geospatial technologies, social perception of place, and some guidance on how to avoid erroneous situations involving Null Island. 

\subsection{Geospatial technology} \label{geospatial}

\begin{figure*}
  \includegraphics[width=\textwidth]{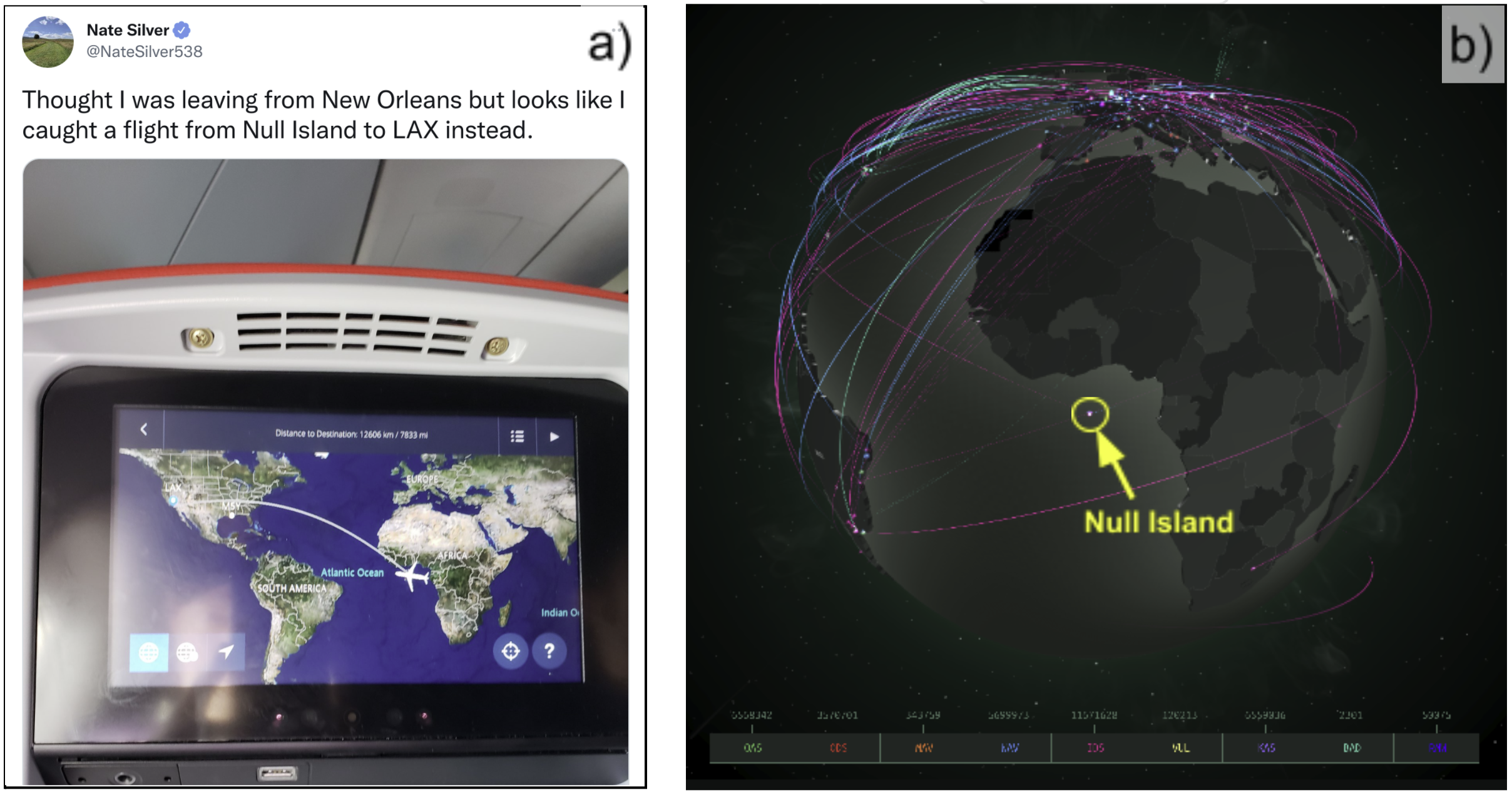}
  \caption{Misleading maps indicating that (a) a flight left from Null Island instead of New Orleans (See original post at \url{https://web.archive.org/web/20220430171046/https://twitter.com/NateSilver538/status/1078748940974084096}), and (b) some cyber attacks originate from Null Island}
  \label{fig:misleading}
\end{figure*}

Although Null Island is a point object, its existence raises several questions and issues across disciplines. These implications together elevate Null Island to becoming a global issue that must be considered in spatial applications and studies. Perhaps the most common implication that concerns the layman is inaccurate visualization when Null Island’s location is present in geographic datasets. According to \cite{monmonier_how_2018}, most ‘blunders that mislead’ are a result of cartographic inattention and inadequate editing. Although these blunders are not intended to purposefully mislead  not all map users are informed or are aware of cartographic fallibility. We found several examples from real-world applications that are incorrect, such as an in-seat map on a flight indicating the take-off location to be Null Island instead of New Orleans, posted by a user on Twitter with over 1.5 million followers (Fig.~\ref{fig:misleading}a). Sharing misleading maps with large audiences is problematic in the era of fake news, since people tend to trust maps as facts, and therefore a misleading map showing Null Island’s location might be misinterpreted by someone, go viral, and even be promoted by the media \cite{mooney_mapping_2020}. The issue is especially pronounced in the context of web maps  that are  are not necessarily made by trained professionals. Another example of this is the ‘rediscovery of Null Island’ by a user on Reddit, who posted an ‘investigation’ with the title \textit{‘I think I discovered a secret Chinese military base in the middle of the ocean’} in a community of 1.5 million members \cite{reddit_i_2021}. In the ‘investigation’, the user referred to several data web maps that showed a ‘mysterious location’ in an island in the middle of the ocean, such as a map showing cyber attacks real time (Fig.~\ref{fig:misleading}b\footnote{ Kaspersky Cyberthreat real-time map: \url{https://cybermap.kaspersky.com/}}), and the existence of an island was also ‘confirmed by’ Strava’s activity heat map (Fig.~\ref{fig:examples}b). 

The implications of our discussion above reach far beyond geovisualization. Other applications, such as geospatial analysis, geocomputation, spatial data storage as well as geospatial programming all experience consequences. In the simplest form, a spatial join of attribute data with an incorrectly placed data point on Null Island can result in poor visualizations, such as the 1896 Summer Olympics medal map by country demonstrated in Fig.~\ref{fig:website}b. With geography’s transformation into a data-driven discipline an increasing number of studies are conducted on large-scale and complex datasets at multiple geographic scales \cite{miller_data-driven_2015}. For example, studying global volunteered geographic information (VGI) or other user-generated datasets can be used to infer the home location of users (citizens) \cite{heikinheimo_detecting_2022}, to predict human mobility \cite{shen_novel_2022} or to analyze global citizen sentiment during the COVID-19 pandemic \cite{okango_dictionary_2022}, to name but a few. Unlike local or regional studies, global datasets have the possibility to include data on Null Island (see Section~\ref{container} for examples). Missing or incomplete data in social media feeds can be overcome by both advanced and manual verification approaches \cite{ilieva_social-media_2018}. It is also routine practice to simply exclude data for other reasons such as bot activity or to mitigate the bias caused by users with tendencies to contribute very little or too much \cite{lovelace_big_2016}. There are no universally adopted guidelines as to how to deal with these cases. However simply excluding data on Null Island and assuming that these events did not occur can be problematic. One problem with removing such events is that it assumes complete randomness to these ''missing'' data points. This is at odds with Table~\ref{tab:sources} which suggests systematic processes are at play leading to locations associated with Null Island. Subsequently, dropping these locations altogether might not be justified statistically. We argued that poorly configured web maps can alter the way users interact with web maps by introducing artificial patterns (Fig.~\ref{fig:tiles}). Therefore, studies analyzing map viewing behavior (see e.g. \cite{mooney_openstreetmap_2021}) should be aware of this and not mistake increased data volume (e.g. at Null Island) with increased activity.

In Wikipedia, currently there are only two articles explicitly assigned the coordinates of (0, 0), namely Null Island and the Gulf of Guinea\footnote{ This can be confirmed by the following MediaWiki API call that searches geotagged articles within a 1 km radius from Null Island: \url{https://en.wikipedia.org/w/api.php?action=query\&list=geosearch&gscoord=0\%7C-0\&gsradius=1000\&gslimit=100}
}. However, as we demonstrated already, (0, 0) is a distinguished place with a significant role, and therefore other articles might also be indexed at the same location by thematic map search engines if they are assumed to be relevant. For example, Frankenplace, a prototype thematic map search engine shows multiple Wikipedia articles on Null Island, such as Earth, Geothermal gradient, Biosphere, and more \cite{adams_frankenplace_2015-1}. This is not an issue itself, however, error propagation in geospatial linked data can be problematic. Referemce \cite{janowicz_moon_2016} demonstrates: \textit{‘Earth therefore can be located at (0,0) together with the statement that its population is 6,814,400,000. [...] Hence, it is the most populated geographic feature in the Gulf of Guinea and thus causes the gulf to have the world’s highest population density.’}

\subsection{The social perception of place} \label{social}

The translation of Null Island’s Wikipedia article into 17 different languages suggests both a growing interest and awareness of this location and that Null Island is transforming into a larger phenomenon reaching beyond technology-oriented communities. In January 2022, a discussion and debate began on the OpenStreetMap main mailing list following the deletion of Null Island’s OSM object with the title ‘Was the deletion of Null Island reasonable?’ \cite{albrecht_was_2022}. In this debate, OSM contributors argued fervently for or against the removal of Null Island from OSM. One side of the debate argued that OSM thrives when OSM map data are verifiable on the ground\footnote{ See \url{https://wiki.openstreetmap.org/wiki/Ground_truth}} and consequently a group of contributors (including members of the authoritative OSM Data Working Group) think that fictional places should not be added to the OSM database. A counter argument shared by many contributors is that many suburbs and localities also do not exist in a physical form. These divisions of geographic space exist as a shared knowledge of locals inhabiting an area. In this regard, Null Island is no more fictional than localities that exist only in the collective consciousness of people, and refers to some specific area or location. This debate within OSM resurfaces from time-to-time and one can guess that there is no apparent resolution on the horizon. The discussion also resembles what is known as the locality debate in the United Kingdom in the 1980s and 1990s to explain the restructuring of economies and their spatial structures. As far back as 1991, it was argued that localities are not simple spatial areas that are defined by an outline, but they should be defined in terms of the sets of social relations or processes in question \cite{massey_political_1991}. One might argue that Null Island fits into this definition of locality, since its concept as well as its ‘real’ location are collectively known by many people, and it is important enough to be regularly part of social discussions in various channels.
 
Although we focus on the technological aspects of Null Island, explaining the growing popularity of it can also be approached from a social perspective. People have always been drawn to geographic oddities and superlatives. Although many of these places have physical properties, for example being the tallest or southernmost points, other locations are purely sentimental in nature such as quadpoints or tripoints. In fact, many geodetic lines and national boundaries became popular tourist attractions because they tend to fascinate people \cite{loytynoja_development_2008-1}. There is also a body of literature developing algorithms to calculate the locations of poles of inaccessibility \cite{barnes_optimal_2020-1, rees_finding_2021}. As \cite{Joliveau10.1179/000870409X415570} argues \textit{'once located on a map, the fictional place becomes attractive for tourists and a potential source of profit'}. In ~\cite{LeePotterdoi:10.1177/1468797612438438} the author asks how value becomes attached to places that are not real but fantastical constructions belonging to the realm of ``secondary worlds (possible non-actual worlds), such as Narnia and Pandora''. The author gives the example of Platform $9\frac{3}{4}$ from Harry Potter which is exemplary of an affective, liminal space. While  Platform $9\frac{3}{4}$ does not exist this does not devalue its meaningfulness and value to tourists who for many the platform at King’s Cross station London has become a reality. The desire to visit an interesting location is not new. An early documented example dates back to Captain James Cook’s expedition to Antarctica in the 1700s. Joseph Bank, a botanist who planned to participate, was left out of the expedition due to a dispute. His biggest disappointment was that he did not get the chance to stand on the South Pole and turn a full circle on his heel through 360\textsuperscript{o} of longitude \cite{timothy_collecting_1998-1}. This attraction is similar to what we see happening in Null Island’s case that actually had been visited multiple times even though there is nothing but vast ocean with a weather observation buoy (Fig.~\ref{fig:where}b) at that location. Among the documented visits are United States Coast Guard vessel Sherman in 2001, project ‘Towards Zero’ in 2007 \cite{degree_confluence_project_0n_2022} and most recently, Russian missile cruiser Marshal Ustinov in 2019 \cite{ministry_of_defence_missile_2019}.

\section{Discussion} \label{discussion}

In 2008, the term Null Island was used for the first time by Steve Pellegrin to describe ‘data goofs’: geographic data that accidentally get assigned to the coordinates of (0, 0) and are then rendered at the origin of the WGS84 geographic coordinate system in the Gulf of Guinea. Since then, Null Island refers to this specific location at the intersection of the equator and the prime meridian. The term has gone through different evolutionary phases, from being used jokingly by the geospatial community to slowly entering mainstream media. Eventually, Null Island has emerged and established itself in GIScience and can now legitimately be considered a fundamental and conceptual issue of geographic information. 

\subsection{Guidelines to avoid Null Island} \label{guidelines}

Here we present practical guidelines to help GIScientists, geospatial professionals and programmers avoid creating, and subsequently having to deal with, issues related to Null Island. Most of the issues we discussed earlier (see Table~\ref{tab:sources}) are associated with errors during programming or geospatial data handling. In the geovisualization context, most of these problems can be avoided by being more attentive and using appropriate techniques during the cartographic process \cite{monmonier_how_2018}. Since mapmaking today is a process involving data handling and programming these issues can originate from geospatial professionals and geoscientists not being experienced in programming and software development. Software developers and programmers not being experienced in understanding geographic principles can also be a contributory factor. To minimize these errors, it is beneficial for geospatial professionals to acquire training in programming and data management  and for programmers who work on geospatial applications to familiarize themselves with basic geographic and cartographic concepts. Perhaps the easiest and most efficient method to spot erroneous data points infiltrating a dataset is visually inspecting datasets on a map. Null Island’s location falls in the Gulf of Guinea in the Atlantic Ocean, roughly 600km off the coast of West Africa. This positioning separates the location easily from most other data points commonly generated over landmasses. Below we list our recommended steps that can be taken to avoid creating issues with Null Island:

\begin{itemize}
 \item Visual inspection of data (maps) for locational correctness
 \item Use of standard software libraries for data input/output data; avoid implementing custom software solutions that can introduce errors
 \item Implement proper exception handling to avoid the null value problem in RDBMS and applications \cite{berztiss_exceptions_2007}
 \item Software testing of custom software components for failure and incorrect behavior \cite{ammann_introduction_2016}
 \item Use of debugging tools, e.g. to visualize errors \cite{keim_information_2005}
 \item Geospatial training for programmers and developers
 \item Technical training (e.g. data management and programming) for geospatial professionals
\end{itemize}

\subsection{A shifting socio-technological concept}
The presence of four evolutionary phases identified in this paper suggests that the concept of Null Island is not static. In the early days, it was purely a technological issue associated with geocoding failures and computerized mapping systems failing to correctly process coordinates. However, its significance is shifting from being a serious technological issue towards one which is more social and even philosophical in nature. Null Island as a social phenomenon can be witnessed for example by branded merchandise for sale (e.g. t-shirts) and social events, memes about Null Island circulating mainly on the internet. The philosophical aspect is present in mapping debates, namely, whether an imaginary place should be part of map databases (e.g. OpenStreetMap). One side of the argument is that maps are a representation of the Earth and imaginary or fictional places should not be part of it. On the other hand, Null Island as a fictional place is no less real than a locality that exists only in the collective consciousness of certain groups of people, but without politically set boundaries. In this sense, Null Island exists as a real place (even though not as a physical island)  since it refers to a specific location on Earth that is known to many people. Null Island is not the first non-existent place going through this transformation, as it resembles closely the story of Agloe, NY, that was originally intended as a copyright trap on paper maps, but became a real locality featuring a general store and gas station (Fig.~\ref{fig:agloe}). Null Island's shift from technological to social concept can be attributed to multiple factors  both technological and social in nature. On the technological side, most online geocoding services (e.g. Google) fixed these initial issues and improved exception handling that originally reverted locations to Null Island when a match was not found. Another reason for Null Island’s decreasing technological significance is that NoSQL solutions are gaining momentum in place of traditional RDBMS and are becoming more common. NoSQL databases are efficient in working with ‘messy’ data such as some of the user-generated sources presented in this paper by not enforcing a fixed database schema \cite{kononenko_mining_2014}. A result of this flexibility is there is no need to store NULL values when pieces of data are missing (e.g. location) which in turn decreases the chances of obtaining (null, null) as a coordinate location from such databases. Even with the changing technology landscape, encountering data on Null Island is still common today as was illustrated in Section~\ref{what_is_null_island}. Even widely popular services operated by multi-billion dollar companies (e.g. Twitter, Snapchat, AirBnB) display data on Null Island. Coupled with the popularization of Null Island by mainstream media outlets has seen it discovered by the general population  and the arts. Humans are traditionally attracted to geographic oddities, superlatives and other extreme points, including imaginary and non-physical features, such as administrative borders, geographic lines and points of inaccessibility. In light of this attraction, it is only natural that Null Island is considered a social phenomenon on top of the clear technology-related nature of it.

\subsection{Moving forward}
We believe that Null Island will stay with us in the long run since programming mistakes are easy to make by distracted programmers and cartographers. This is even more pronounced when the programmer is unaware of geographic principles such as map projections or when a geospatial professional is inexperienced in programming. These instances will likely happen over and over again as can be illustrated with the steady supply of Null Island related posts in programming Q\&A sites (e.g. GIS StackOverflow) and social news aggregators, such as the re-discovery of Null Island by a reddit user (see the title and Section~\ref{geospatial}). Null Island may seem like a lighthearted topic that is part of jokes and funny conversations. However misplaced data can have serious implications and therefore it is a topic of interest to GIScience and related disciplines. Implications can include, among others, spreading misleading information through maps, introducing another source of bias into already noisy user-generated datasets, or error propagation in linked geodata. To mitigate these instances, it is advisable to pay extra attention in situations that would introduce Null Island to a dataset. Moving forward we see two opportunities that are sometimes missed but could be easily implemented. First, more technical training in terms of data management as well as programming should be introduced into geospatial technology education so that the new generation of geospatial professionals are fully equipped with skills that are needed in today’s technology dependent environment. On the other hand concerning developers and programmers, it is unreasonable to expect that geospatial training will be part of the mainstream computer science curricula. Perhaps it is the task of the GIScience community to develop accessible crash courses or offer training about geospatial principles, such as projections and coordinate systems to computer science professionals.

\section{Conclusions and future work} \label{summary}

This paper has discussed Null Island, a fictional place located at (0, 0) in the WGS84 geographic coordinate system at the intersection of the equator and prime meridian. It is traditionally used as a placeholder for ‘bad’ and misplaced geographic data in databases. We have explained the importance of discussing Null Island as a conceptual and fundamental issue of geographic information. The are several important contributions from this paper. To our knowledge, after extensive searches, this is the first serious scientific article dedicated to Null Island and the widespread implications of its existence. Our paper has delivered a number of important contributions. The paper provides the first comprehensive treatment of Null Island as a geospatial concept which will be of interest to those in the GIScience. This structured and serious treatment of Null Island will encourage other researchers to consider the broader implications of this real fictional place. We have also presented error sources and practices that link or define geographic data with Null Island. By outlining four evolutionary phases, our paper provides an accurate and evidence-based account on how Null Island is used today. From a practical viewpoint the implications of wrongly attributing data to Null Island was described and the interplay with geospatial location descriptions, spatial relation terms and referencing geographic objects. Guidelines to avoid erroneously placing data on Null Island are also provided. 

Null island will continue to be a shifting socio-technological concept into the future and this paper delivers the first comprehensive treatment of Null Island for the GIScience community. We believe that issues associated with Null Island will continue to be present for mainly two reasons: 1) programming and cartographic mistakes when handling geographic data are common among programmers not familiar with geospatial concepts, and among geospatial professional not familiar with programming, and 2) humans will continue to be intrigued by geographic oddities and interesting places. These two reasons will likely keep Null Island related discussions circulating for some time. This paper is aimed at both GIScientist, programmers and the general population to raise awareness of the implications that are inherently present in the existence of Null Island. The paper can also serve the role of historical record for Null Island for multidisciplinary research work considering exploring this topic further. For future work, we plan to explore the concept of Null Island as a place and gain more understanding into why humans, mappers and geographers are so drawn to geographic oddities like Null Island. Indeed, artistic representations of Null Island open up a larger discussion about fiction, geography and simulated realities which are currently beyond the scope of this paper.

\bibliographystyle{IEEEtran}
\bibliography{IEEEabrv,null_island_access.bib}

\begin{IEEEbiography}[{\includegraphics[width=1in,height=1.25in,clip,keepaspectratio]{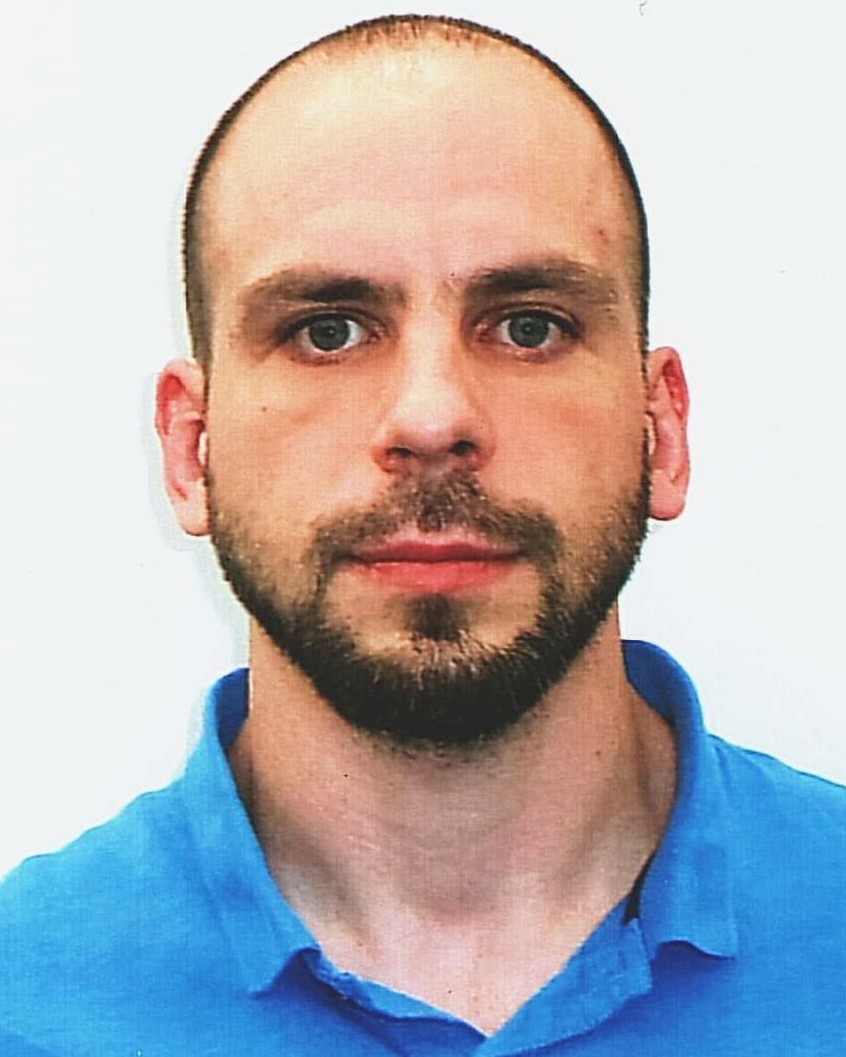}}]{Levente Juhász} (M'2020) received the BSc and MSc degrees in geography with geoinformatics specialization from the University of Szeged, Hungary in 2011 and 2013 respectively, and the PhD degree in Geomatics from the University of Florida in 2018.

His was a Research Assistant at the Department of Physical Geography and Geoinformatics at the University of Szeged, Hungary, and at University of Florida's Geomatics Program at Fort Lauderdale Research \& Education Center in Fort Lauderdale, FL, USA. He has held visiting positions at the Digital Earth and Reference Data Unit, Joint Research Centre of the European Commission in Ispra, Italy and at the Department of Geoinformation Technologies, Carinthia University of Applied Sciences in Villach, Austria. Since 2019, he is research faculty at Florida International University in Miami, FL. Currently, he is a Research Assistant Professor and he serves as the Interim Director of the Geographic Information Systems (GIS) Center. He has authored over 30 peer-reviewed journal articles, book chapters and conference papers. His research revolves around understanding the quality of user-generated geographic information, and is heavily interconnected with technological advancements in geographic information science.
\end{IEEEbiography}

\begin{IEEEbiography}[{\includegraphics[width=1in,height=1.25in,clip,keepaspectratio]{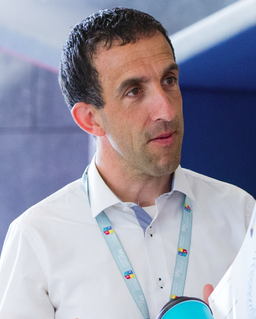}}]{Peter Mooney} is an Assistant Professor at the Department of Computer Science in Maynooth University and has been working in the domain of geospatial data research for over a decade. He is particularly interested in understanding the processes behind the collection of volunteered geographic information data and how computing techniques such as machine learning can be applied to these datasets and sources. He is heavily involved in OSGeo (Open Source Geospatial Foundation) activities in Ireland. His teaching philosophy is one which embraces open science and he uses FOSS4G (Free and Open Source Software for Geomatics) and Open Data for all student teaching and learning activities. He is currently an editor of the Transactions in GIS journal. He has authored over 70 peer-reviewed journal articles, book chapters and conference papers. 
\end{IEEEbiography}

\appendix

\section{Additional tables and figures}

\begin{table*}
\caption{References to additional events associated with Null Island }
\begin{tabular}{l | l | p{4in}} 
 \textbf{Date} & \textbf{Event} & \textbf{Source} \\ \hline
 4/17/2009 & First mention on Twitter & \url{http://web.archive.org/web/20180910184441/https://twitter.com/chriscurrie/status/1546199025} \\    \hline
 11/20/2010 & Outline traced based on Myst Island & \url{http://web.archive.org/web/20170325024549/https://github.com/geoiq/acetate-styles/commit/136d23facbf80953ccb3eb4419a87b2b6ee0bb1a } \\  \hline
  May  2011 & First dedicated Twitter account & \url{http://web.archive.org/web/20140326105433/https://twitter.com/NullIsland} \\ \hline
  10/25/2012 & First comment on Hacker News & \url{http://web.archive.org/web/20121029121942/https://news.ycombinator.com/item?id=4697543} \\ \hline
  10/1/2013 & English Wikipedia entry created & \url{https://en.wikipedia.org/w/index.php?title=Null_Island&oldid=575288934 } \\ \hline 
  5/17/2014 & Wikidata entry created & \url{https://www.wikidata.org/w/index.php?title=Q16896007&oldid=130397167 } \\ \hline
  2014 & Added to Foursquare & \url{http://web.archive.org/web/20220411153347/https://foursquare.com/v/null-island/508c6298e4b019412ca444cf} \\ \hline
  3/28/2014 & First mention on Reddit & \url{http://web.archive.org/web/20220411153501/https://www.reddit.com/r/ProgrammerHumor/comments/21kije/i_just_flew_in_from_null_island_and_boy_are_my/} \\ \hline
  8/13/2014 & Added to Geocommons & \url{https://raw.githubusercontent.com/geoiq/gc_data/master/datasets/104581.geojson } \\ \hline 
  7/15/2016 & First story on Slashdot & \url{http://web.archive.org/web/20211206062041/https://developers.slashdot.org/story/16/07/15/064248/null-island-the-land-of-lousy-directional-data} \\ \hline
\end{tabular}
\label{tab:first_events}
\end{table*}

\begin{figure*}
  \includegraphics[width=\textwidth]{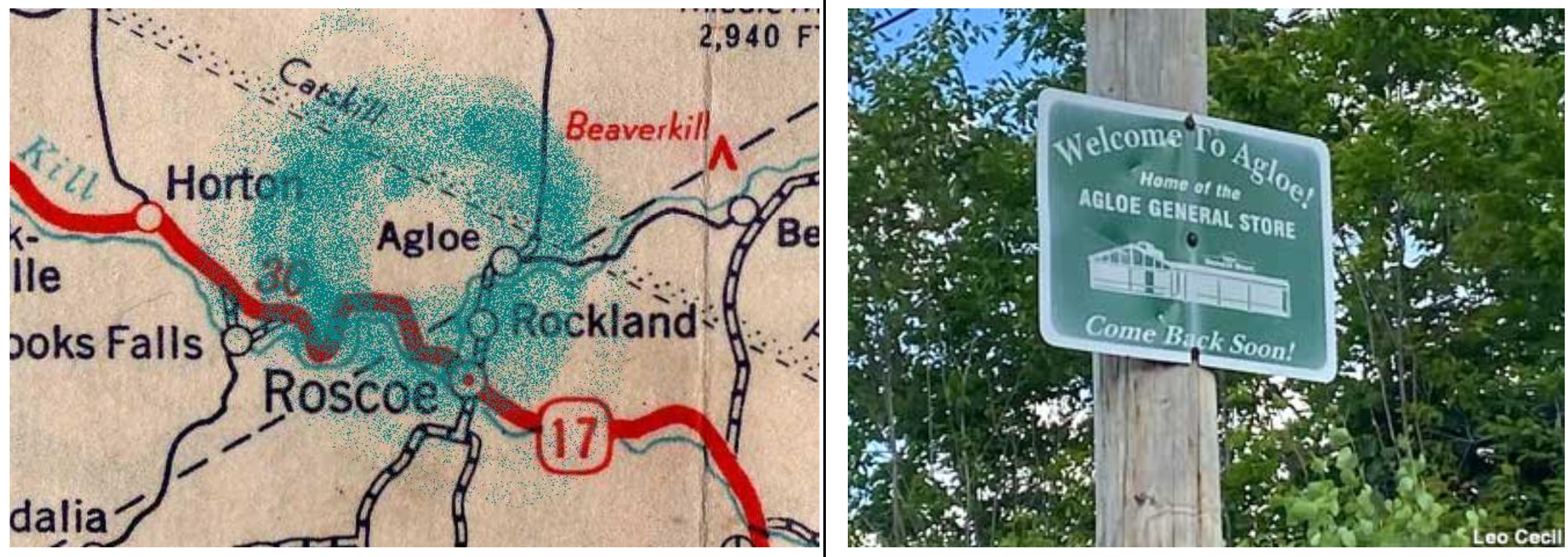}
  \caption{A similar place transformation to that of Null Island: from being a fictious copyright trap in paper maps, Agloe, NY became a real place}
  \label{fig:agloe}
\end{figure*}

\begin{figure*}
  \includegraphics[width=\textwidth]{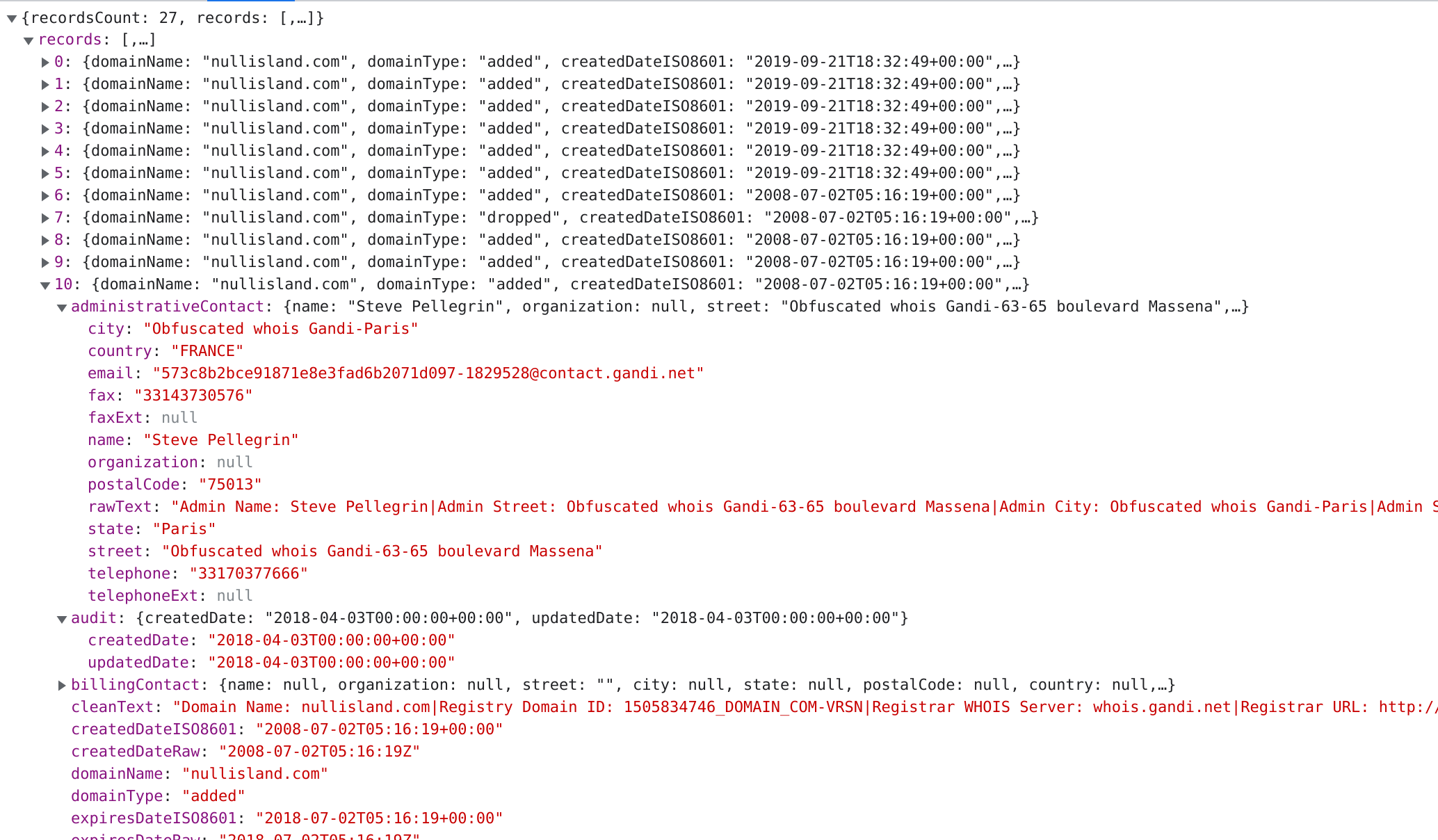}
  \caption{Historical WHOIS records of the domain nullisland.com (Source: \url{https://whois-history.whoisxmlapi.com/}) }
  \label{fig:whois}
\end{figure*}

\begin{figure*}
    \centering
  \includegraphics[width={4in}]{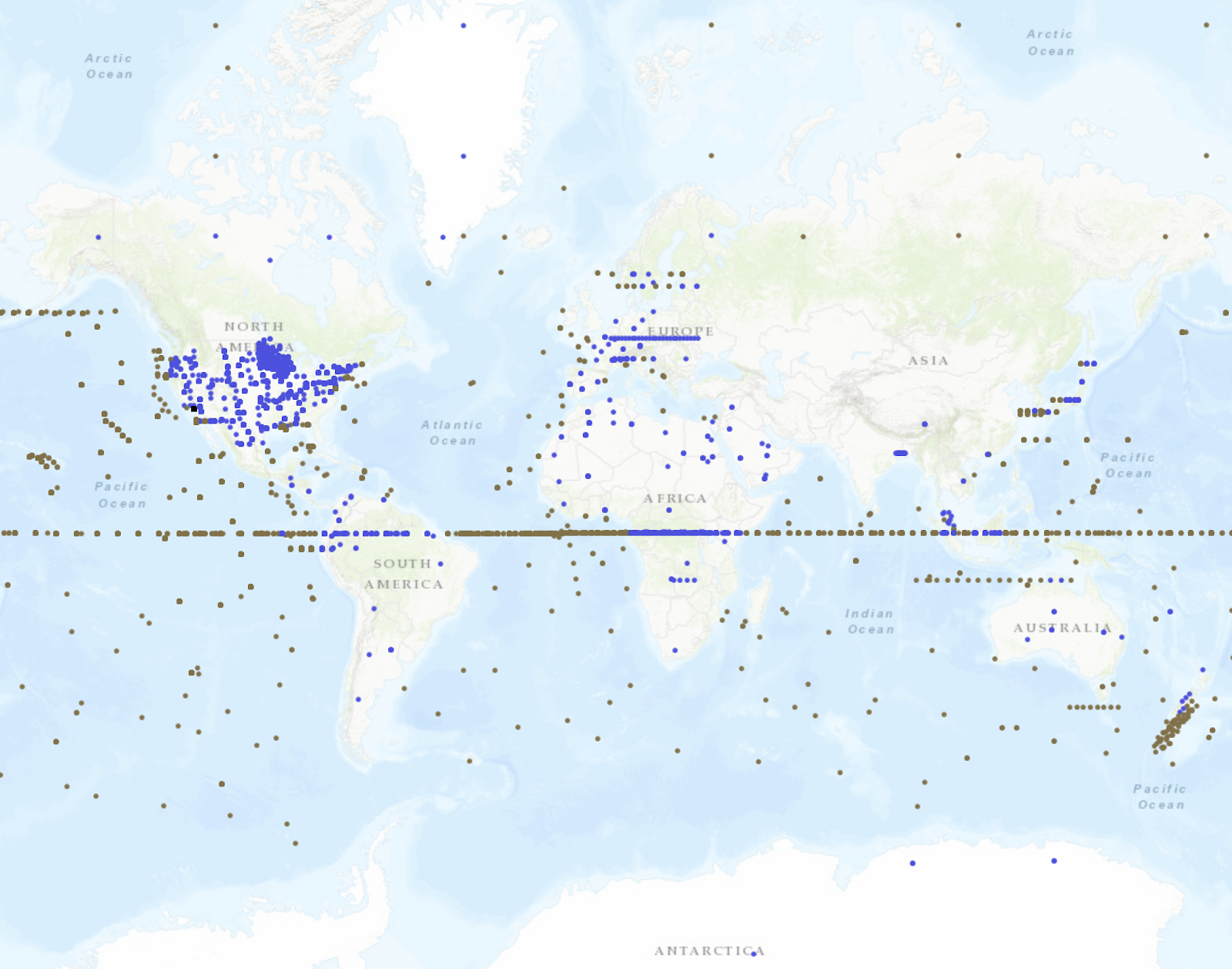}
  \caption{Nill Points story map shows the origins of all coordinate systems supported by ESRI software \cite{field_nill_2014} }
  \label{fig:nillponts}
\end{figure*}

\begin{figure*}
  \centering
  \includegraphics[width={4in}]{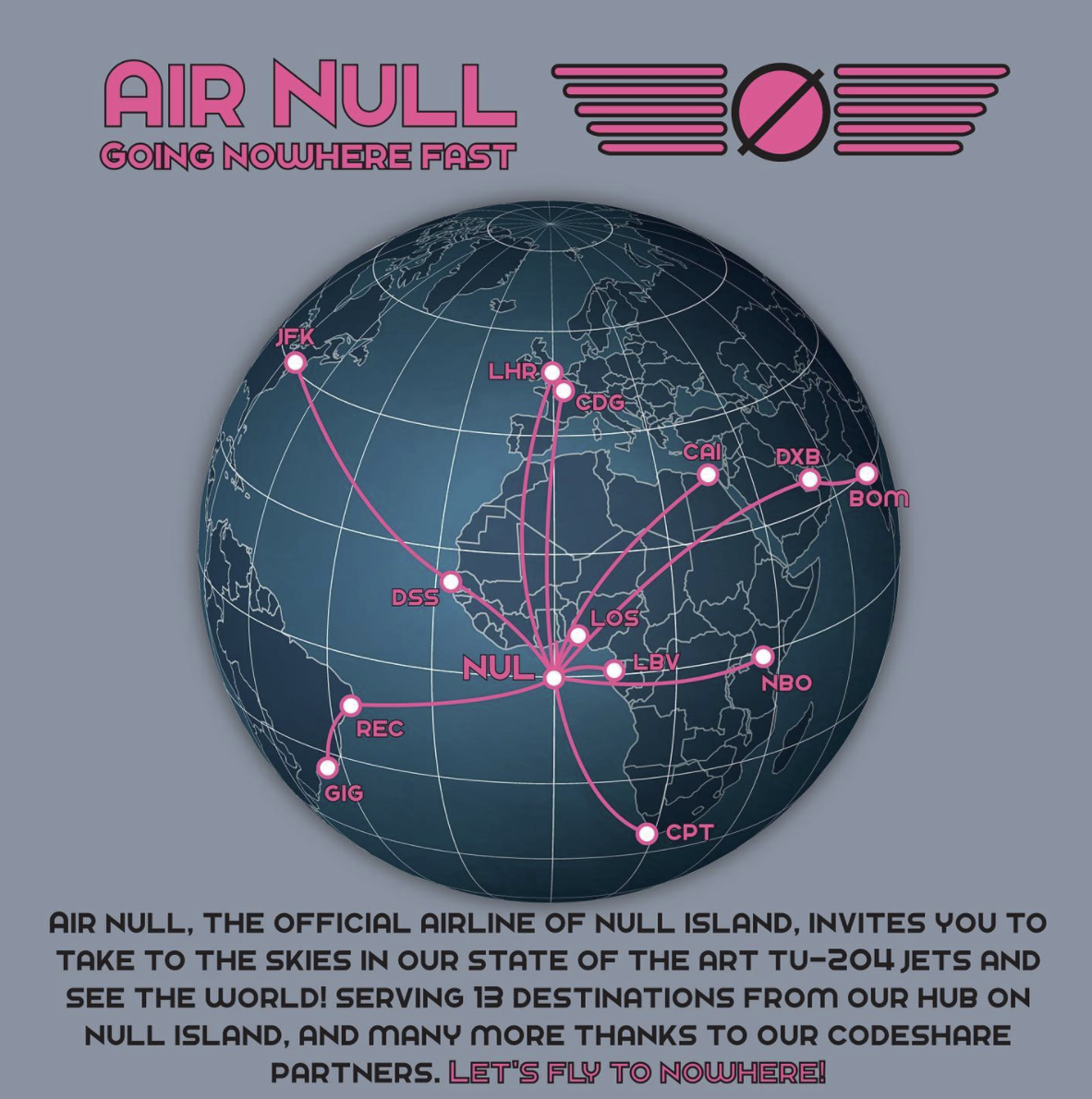}
  \caption{Intentional use of Null Island as a joke (Source: \url{https://www.instagram.com/p/CW235ZqBqzy}) }
  \label{fig:airnull}
\end{figure*}

\begin{figure*}
  \includegraphics[width=\textwidth]{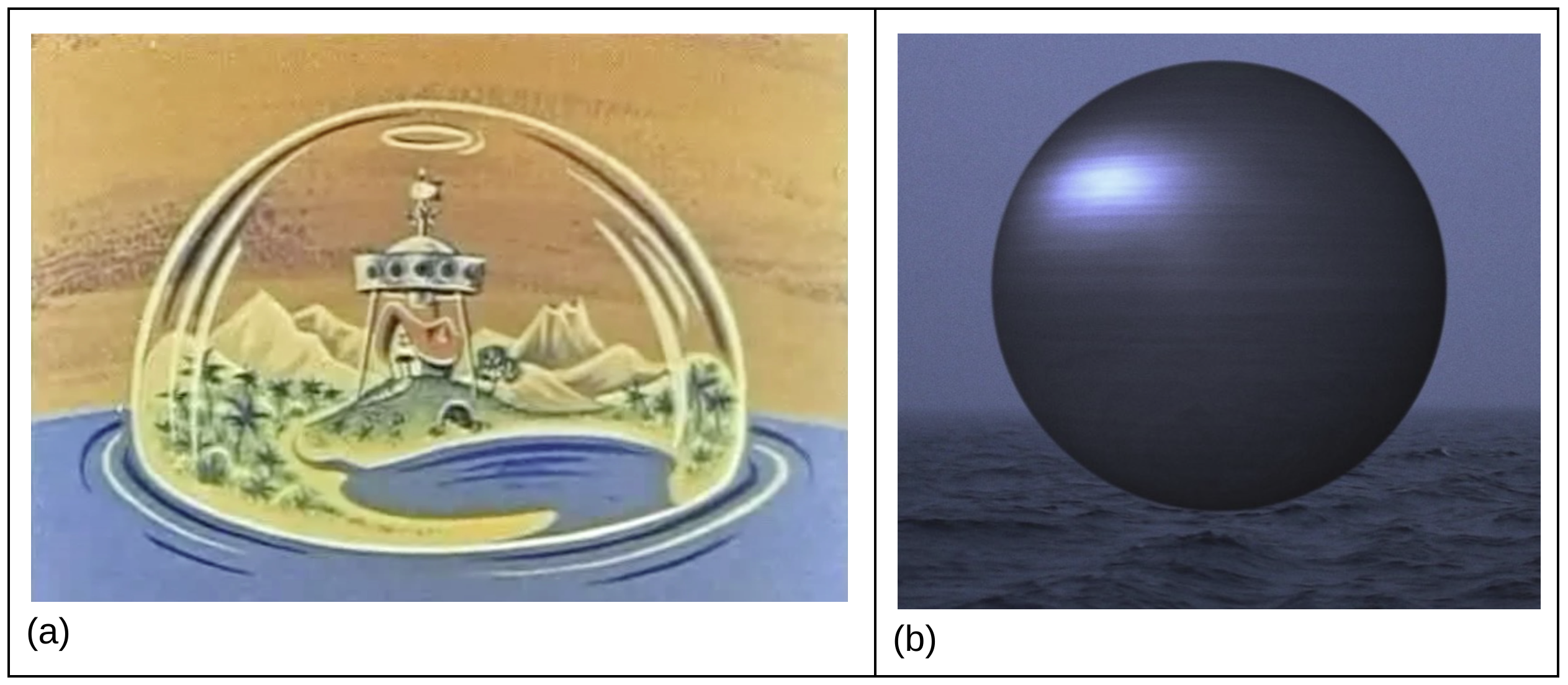}
  \caption{Visual arts inspired by Null Island: (a) Zero Zero Island in Colonel Bleep, the first color cartoon ever made for television in 1957, and (b) Screen grab from an interactive movie featuring Null Island as created by artist Letícia Ramos in 2020 }
  \label{fig:visual_arts}
\end{figure*}

\end{document}